\documentclass[12pt]{article}
\pdfoutput=1
\usepackage{amssymb}
\usepackage{amsmath,amsfonts,feynmp}
\usepackage{graphicx}
\usepackage{tipa}
\usepackage{stmaryrd} 
\usepackage[pdftex, bookmarks=true,colorlinks,linkcolor=red,urlcolor=blue,citecolor=blue]{hyperref}

\makeatletter\@addtoreset{equation}{section}\makeatother

\makeatletter\@addtoreset{equation}{section}\makeatother

\newcommand{\preprint}[1]{\begin{table}[t]  %%
             \begin{flushright}               %%
             {#1}                             %%
             \end{flushright}                 %%
             \end{table}}                     %%
\renewcommand{\title}[1]{\vbox{\center\LARGE{#1}}\vspace{5mm}}
\renewcommand{\author}[1]{\vbox{\center#1}\vspace{5mm}}
\newcommand{\address}[1]{\vbox{\center\em#1}}

\begin{document}

\unitlength = .8mm

\begin{titlepage}
\vspace{.5cm}
\preprint{IFT-UAM/CSIC-14-105}
 
\begin{center}
\hfill \\
\hfill \\
\vskip 1cm

\title{Negative magnetoresistivity in chiral fluids and holography}

\vskip 0.5cm
 {Karl Landsteiner}\footnote{Email: {\tt karl.landsteiner@uam.es}}, 
 {Yan Liu}\footnote{Email: {\tt yan.liu@csic.es} } and 
 {Ya-Wen Sun}\footnote{Email: {\tt yawen.sun@csic.es}}
 
\address{Instituto de F{\'\i}sica Te\'orica UAM/CSIC, C/ Nicolas Cabrera 13-15,\\
Universidad Aut\'onoma de Madrid, Cantoblanco, 28049 Madrid, Spain}

\end{center}

\vskip 1.2cm

\abstract{ In four dimensions Weyl fermions possess a chiral anomaly which leads to several special features in the transport phenomena, such as the {\em negative longitudinal magnetoresistivity}. In this paper, we study its inverse, the longitudinal magnetoconductivity, in the case of a chiral anomalous system with a background magnetic field $B$ using the linear response method in the hydrodynamic limit and from holography. Our hydrodynamic results show that in general we need to have energy, momentum and charge dissipations to get a finite DC longitudinal magnetoconductivity due to the existence of the chiral anomaly. Applying the formula that we get from hydrodynamics to the holographic system in the probe limit, we find that the result in the hydrodynamic regime matches that calculated from holography via Kubo formula. The holographic result shows that in an intermediate regime of $B$ there is naturally a negative magnetoresistivity which decreases as $1/B$. At small $B$ direct calculations in the holographic system suggest that holography provides a new explanation for the small $B$ positive magnetoresistivity behavior seen in experiment, i.e. the small $B$ behavior comes from the quantum critical conductivity being affected by the chiral anomaly.}

\vfill

\end{titlepage}

%\eject \tableofcontents%

%%%%%%%%%%%%%%%%%%%%%%%%%%%%%%%%%%%%%%%%%%%%%%%%
\section{Introduction}
%%%%%%%%%%%%%%%%%%%%%%%%%%%%%%%%%%%%%%%%%%%%%%%%

During the last few years the study of the dynamics of fluids of chiral fermions has received considerable attention.
Apart from purely theoretical interest the theory of chiral fluids might find application in very different physical systems, such
as the quark gluon plasma \cite{Fukushima:2008xe, Kharzeev:2007jp}, advanced materials such as Weyl semi-metals \cite{qi-review} or Proto-Neutron stars \cite{Kaminski:2014jda}.

Probably the most striking property of chiral fermions is the breaking of a classical symmetry via the chiral anomaly \cite{Adler:1969gk,Bell:1969ts}. Although the notion
of fluid is an intrinsically macroscopic notion it does inherit the anomaly from its microscopic origin \cite{Son:2009tf}. It is by now well understood
that the chiral anomaly gives rise to a variety of parity odd and dissipationless transport phenomena, such as the chiral magnetic and
the chiral vortical effects \cite{Erdmenger:2008rm, Banerjee:2008th, Neiman:2010zi,Landsteiner:2011cp, Landsteiner:2011iq, Jensen:2012jy,Jensen:2012kj,Golkar:2012kb,Hou:2012xg,Banerjee:2012iz,Kharzeev:2013ffa}.

Besides these new dissipationless and parity odd transport, the anomaly also has a profound impact on the electric DC conductivity.
More precisely, in a magnetic background field the (longitudinal) electric DC conductivity is strongly enhanced by the magnetic field due to the anomaly. This
has been pointed out first in \cite{Nielsen:1983rb} and more recent studies \cite{sonspivak,gorbar}  have confirmed this idea.
This phenomenon is called {\em negative magnetoresistivity} and should be realized in condensed matter systems such as Dirac- or Weyl (semi-)metals.\footnote{For normal metals without anomaly the magnetoconductivity is monotonically non-increasing with the magnetic field \cite{theorem-prb}.} In %the 
holography this has also been found in the context of the Sakai-Sugimoto for the baryon DC conductivity model
in \cite{Lifschytz:2009si}.\footnote{We also remark that in QCD in the confined phase a strong magnetic enhancement of the electric conductivity has been found
in \cite{Buividovich:2010tn,{Buividovich:2010qe}} In fact for large enough magnetic field the QCD vacuum becomes even an anisotropic superconductor \cite{Chernodub:2010qx}.}
As we will show, this phenomenon can be derived from the form of anomalous hydrodynamics using linear response theory.

A recent experiment \cite{weyl-exp} in which a magnetic field was applied to Bi$_x$Sb$_{1-x}$ at $x\sim 0.3$ found
indeed a negative magnetoresistivity behavior in an intermediate regime of the magnetic field $B$. The negative magnetoresistivity behavior could be calculated in the large $B$ limit using weakly coupled theoretical methods \cite{{Nielsen:1983rb},{sonspivak}}, which decreases as $1/B$. In the experiment, at small $B$, there is an increase in the magnetoresistivity as a function of $B$, which has a different origin from the negative magnetoresistivity in an intermediate regime of $B$. The dependence of the magnetoconductivity on the magnetic field in the whole range of $B$ is not easy to calculate theoretically even in the weak coupling limit and the small $B$ behavior from weakly coupled field theoretical calculations does not fit the experimental data well. In \cite{{weyl-exp},{kim-boltzman},{kim-review}}, the authors tried to explain this behavior by the weak anti-localization effect, which is a phenomenological model incorporating weak anti-localization quantum corrections. All these are weakly coupled results, and in this paper we would like to use AdS/CFT correspondence to study the longitudinal magnetoresistivity of a strongly coupled chiral fluid, especially its dependence on the background magnetic field. As AdS/CFT is a tool to use weakly coupled gravity to study the strongly coupled field theory, our results will be related to strongly coupled physics, and this will produce similar small $B$ behavior as in experiments but with a different origin.

Before starting the holographic calculations of the magnetoconductivity of a chiral fluid, recall that in a translationally invariant system and in the absence of any mechanism of momentum dissipation, there will be an infinite DC conductivity as the charge carriers can accelerate to infinite momentum under an external electric field at zero frequency. For a charged chiral fluid things are more complicated due to chiral anomaly. Without chiral anomaly the handedness of the Weyl fermions is conserved. The chiral anomaly  will however induce a  transfer of charge density between the left- and right-handed Weyl fermions and this in turn will result in an anomaly related infinite longitudinal DC magnetoconductivity.
For the case of ordinary metals, the infinite DC conductivity caused by translational invariance can be relaxed by momentum dissipation terms. For the case of longitudinal magnetoconductivity  we will study the dissipation effects in the transport behavior of a chiral anomalous system to see if this is still the case. We will employ the linear response method in the hydrodynamic limit and turn on all the possible dissipation terms, i.e. charge relaxation, momentum relaxation and energy relaxation. Our result shows that in general all the dissipation terms are needed to get a finite longitudinal DC magnetoconductivity. In certain limits, only one or two of them are needed, such as the zero density limit, where only charge dissipation is needed \cite{Jimenez-Alba:2014iia}. In the context of Weyl metals the origin
of charge dissipation can be traced back to the finite inter-valley scattering time. Indeed the chiral (or better axial) symmetry is only
an emergent or accidental one and is broken by tree level coupling akin to a mass term in the Dirac equation. Momentum dissipation due to disorder is of course a generic property of the electron gas in a metal, our result suggests however that also inelastic processes leading to
energy relaxation play an important role. 

Using the linear response method we get a formula for the longitudinal magnetoconductivity in the hydrodynamical limit, whose form does not rely on the microscopic details. Applying this formula to the system dual to AdS Schwarzschild black hole, we will find that the result in the hydrodynamic regime is exactly the same as the result obtained from holographic calculations via Kubo formula. In the holographic result, we will be able to reproduce the negative magnetoresistivity, and the strongly coupled result from holography coincides with the weakly coupled results in \cite{Nielsen:1983rb,{sonspivak},{gorbar}}, as well as the experimental data in an intermediate regime of $B$, which exhibits a $1/B$ negative magnetoresistivity behavior. In the small magnetic field limit, direct calculations from holography give a new explanation for the small $B$ positive magnetoresistivity behavior found in experiment.

The rest of the paper is organized as follows. We will first calculate the electric conductivity of a chiral anomalous fluid with a background magnetic field in the hydrodynamic limit using the linear response method in section \ref{sec2} with all possible dissipation terms turned on and obtain a formula for the longitudinal conductivity. In section \ref{secapply}, we apply this formula to the chiral anomalous system dual to the AdS Schwarzschild black hole in the probe limit of the gauge field. In section \ref{sec3} we will directly calculate the same magnetoconductivity for the holographic chiral anomalous system in the probe limit via Kubo formula and show that the result matches the hydrodynamic formula in this limit. The behavior of the magnetoconductivity as a function of $B$ can also be obtained after assuming an appropriate value of the charge relaxation time. Then we will generalize these calculations to the $U(1)_V\times U(1)_A$ case in section \ref{sec4} which is a more natural case with a conserved electric current besides the anomalous axial current. Discussion of these results and open questions will be presented in section \ref{sec-con}.

%%%%%%%%%%%%%%%%%%%%%%%%%%%%%%%%%%%%%%%%%%%%%%%%
\section{Magnetoconductivity from chiral anomalous hydrodynamics}
\label{sec2}
%%%%%%%%%%%%%%%%%%%%%%%%%%%%%%%%%%%%%%%%%%%%%%%%

Due to the chiral anomaly 
\begin{equation}
\partial_\mu J^\mu=cE^\mu B_\mu
\end{equation}
a background magnetic field will induce a large longitudinal DC conductivity \cite{Nielsen:1983rb,{sonspivak},{gorbar}}.
In particular the axial anomaly turns left (right) handed fermions into right (left) handed ones and after infinite time (and
in the absence of tree level breaking of axial charge)
there will be an infinite (axial) chemical potential due to the chiral anomaly. This results in an infinite DC conductivity in the direction  along the background magnetic field. The mechanism can be understood in more detail as follows.

\begin{figure}[h!]
\begin{center}
\begin{tabular}{cc}
\includegraphics[width=0.95\textwidth]{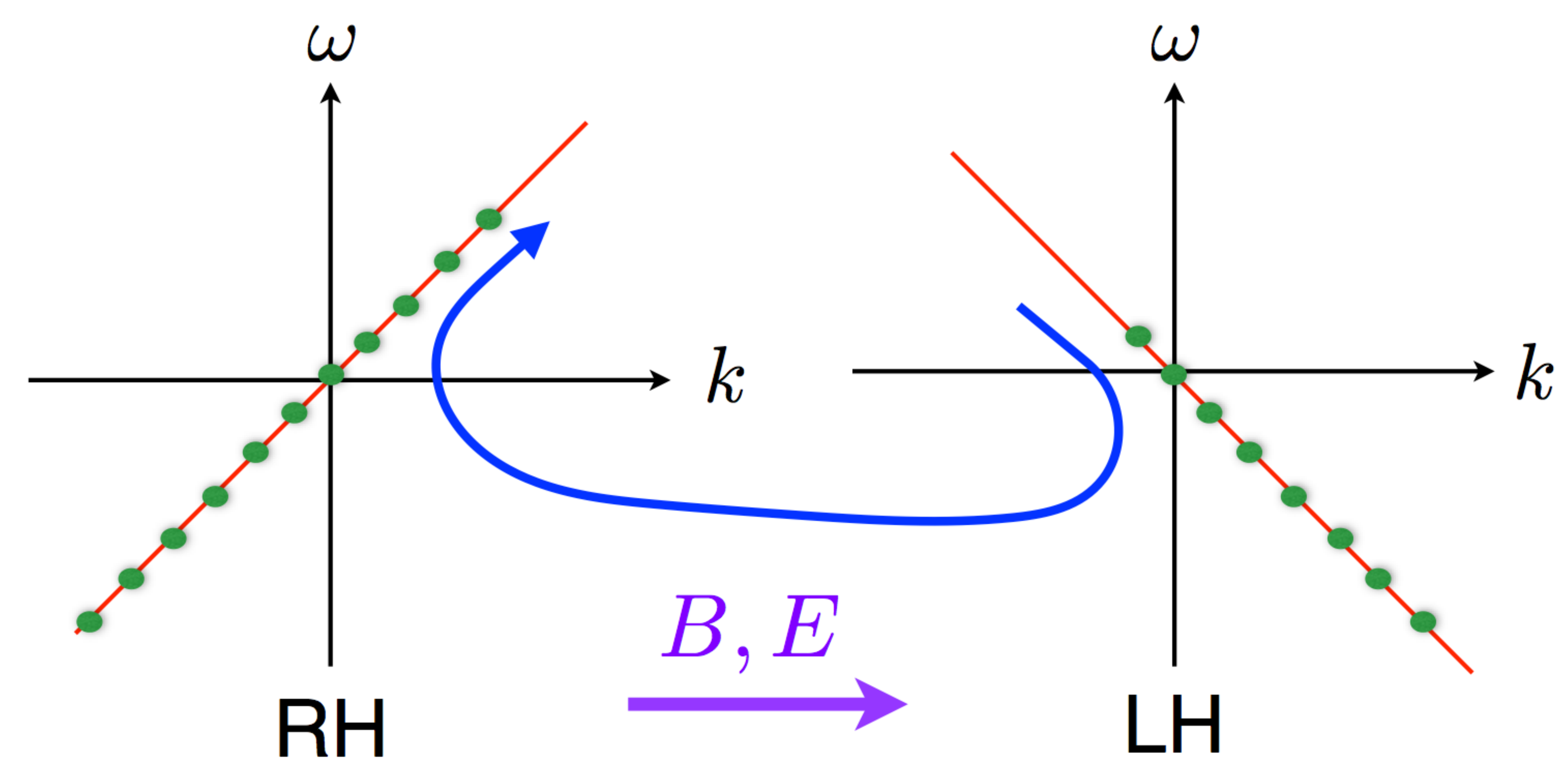}
\end{tabular}
\end{center}
\caption{\small The picture shows the charge transfer under parallel external magnetic and electric fields for Weyl fermions with positive electric charge. Weyl fermions transfer from the left handed band to the right handed one, which is caused due to the chiral anomaly.}
\label{cartoon}
\end{figure}

 Let us consider a system of Weyl fermions with positive charge for both the right handed and left handed fermions. When there is a background magnetic field pointing in the $z$ direction the spectrum will organize into Landau Levels. Only the lowest Landau level is relevant for
 our discussion.
 The spins of the Weyl fermions in the lowest Landau level all point in the $z$ direction. For the right handed particles, their momentum will also point in the $z$ direction while for the left handed ones the momentum will point in the $-z$ direction. Now if we add a small external electric field in the same direction as the magnetic field, the Weyl fermions will accelerate in this direction and left handed fermions will turn into right handed ones. This mechanism is effective even in the zero density limit. At zero density, i.e. in the vacuum, when we add an extra external electric field pointing in the same $z$ direction, the antiparticle of the right handed fermions, which are left handed with negative electric charge will point to the $z$ direction, and they will accelerate in the $-z$ direction under the external electric field which will soon turn them into right handed Weyl fermions with negative electric charge. This means that we now excited a positive chemical potential for the right handed positive charges and a negative chemical potential for the left handed positive charge (which corresponds to a positive chemical potential of right handed negative charges). One can also say that starting from the vacuum and a magnetic field
an additional electric field will induce an axial chemical potential via the anomaly. This in turn will trigger the chiral magnetic effect.
Since (without dissipation) the axial charge will grow without bound the DC conductivity will end up being infinite.
This can be easily understood in the following picture Fig. \ref{cartoon}.

%%%%%%%%%%%%%%%%%%%%%%%%%%%%%%%
%%%%%%%%%%%%%%%%%%%%%%%%%%%%%%%%%%%%

%%%%%%%%%%%%%%%%%%%%%
\subsection{Linear response}
%%%%%%%%%%%%%%%%%%%%%

Here our motivation is to consider the effects of different dissipation terms in the longitudinal DC conductivity. We start from a more universal setup and directly calculate the longitudinal DC conductivity in the hydrodynamic limit in four dimensions with a background magnetic field at the linear response level. Using the linear response method in hydrodynamics developed in \cite{linearresponse}, we can get a result which does not rely on the underlying microscopic details and from our results we will see that without any dissipation terms, there will be several infinite contributions to the longitudinal DC conductivity and different dissipation terms are needed to make it finite.

The linear response method was also used in \cite{nernsteffect} to obtain the magnetoconductivity in 2+1 dimensions. The procedure is to first perturb a hydrodynamic system in a given equilibrium state and solve the system with initial values of the perturbations, then various transport coefficients can be obtained from the response of the electric or thermal current to the initial values of corresponding perturbations.

To perform this procedure and obtain the electric conductivity we will first write out the hydrodynamic equations for the four dimensional chiral anomalous fluid. In this section we focus on the simplest case of only one $U(1)$ current with a triangle anomaly. The conservation equations for the energy momentum and current of a chiral anomalous fluid are \cite{Son:2009tf}
\begin{eqnarray}
\partial_\mu T^{\mu\nu}&=&F^{\nu\alpha}J_{\alpha}\,,\\
\partial_\mu J^\mu&=&c E^\mu B_\mu\,,
\end{eqnarray}
where $c$ is the anomaly constant and $E^{\mu}$, $B^{\mu}$ are background electric and magnetic fields. The corresponding constitutive equations for $T^{\mu\nu}$ and $J^\mu$ are
\begin{eqnarray}
T^{\mu\nu}&=&(\epsilon+p) u^\mu u^\nu +p g^{\mu\nu}+\tau^{\mu\nu}\,,
\\
J^{\mu}&=& \rho u^\mu+ \nu^\mu\,,
\end{eqnarray}
where $\tau^{\mu\nu}$ and $\nu^{\mu}$ are first order corrections in the derivative of hydrodynamic variables, $\rho$ is the charge density and $u^{\mu}$ is the local fluid velocity which satisfies $u^{\mu}u_{\mu}=-1$. In Landau frame the most general forms of $\tau^{\mu\nu}$ and $\nu^{\mu}$ are \cite{{Son:2009tf},Neiman:2010zi}\footnote{Our convention is $g_{\mu\nu}=\text{diag}(-1,1,1,1)$ and $\epsilon_{0123}=-1.$}
\begin{eqnarray}
\tau^{\mu\nu}&=&-\eta P^{\mu\alpha} P^{\nu\beta}(\partial_\alpha u_\beta+\partial_\beta u_\alpha)-(\zeta-\frac{2}{3}\eta)P^{\mu\nu}\partial_\alpha u^\alpha\,,\\
\nu^\mu&=&-\sigma_E TP^{\mu\nu}\partial_\nu\big(\frac{\mu}{T}\big)+\sigma_E E^\mu+\sigma_V \omega^\mu+\sigma_B B^\mu\,,
\end{eqnarray}
where $P^{\mu\nu}=g^{\mu\nu}+u^{\mu}u^{\nu},$ 
\begin{equation}
E^\mu=F^{\mu\nu}u_\nu,~~~ B^\mu=\frac{1}{2}\epsilon^{\mu\nu\alpha\beta}u_\nu F_{\alpha\beta},~~~\omega^\mu=\frac{1}{2}\epsilon^{\mu\nu\alpha\beta}u_\nu\partial_\alpha u_\beta
\end{equation}and
\begin{eqnarray}
\sigma_B&=&c \mu-\frac{1}{2}\frac{\rho}{\epsilon+p}(c\mu^2+c_gT^2)\,, \nonumber\\ 
\sigma_V&=& c\mu^2+c_g T^2-\frac{2\rho}{\epsilon+p}\big(\frac{c\mu^3}{3}+c_g\mu T^2\big)
\end{eqnarray}
with $c_g$ related to the gravitational anomaly. 
  $\sigma_B$ is the chiral magnetic conductivity and $\sigma_V$ is the chiral vortical conductivity. These two terms arise due to the effects of the chiral anomaly. For lack of a better name we call $\sigma_E$ the quantum critical conductivity following \cite{nernsteffect}. It may also depend on the chemical potential or temperature, but its explicit form is not universal and therefore cannot be uniquely fixed from hydrodynamics. In the hydrodynamic regime, we assume that $T\geq \mu$ and $E, B\ll T^2$ so that the first derivative expansions are the leading contributions in $\nu^\mu$ and we can ignore higher derivative expansion terms,\footnote{For hydrodynamic of larger $B$ case, see e.g. \cite{Hansen:2008tq,{geracie-son}} for 2+1 dimensional case.} and to be careful enough we also assume that $|c B|\ll T^2$ as in the first derivative expansion in $J^{\mu}$, $B$ enters in the combination of $c B$ where $c$ is a dimensionless number that can be either large or small.

We assume that the system is in an equilibrium state in the grand canonical ensemble with chemical potential $\mu$, temperature $T$ and the local velocity $u^t=1$. Other thermodynamic variables are determined by these variables and satisfy
\begin{equation}
\epsilon+p=Ts+\mu\rho\,,~~~dp=sdT+\rho d\mu\,.
\end{equation}
The nonzero components of $T^{\mu\nu}$ and $J^\mu$ are
\begin{equation}
T^{00}=\epsilon\,, ~T^{ii}=p\,, ~J^t=\rho\,,~J^z=\sigma_B B\,.
\end{equation}

To calculate the electric conductivity of this system with anomalous effect turned on, we assume that there is a background magnetic field in the $z$ direction which without loss of generality we take to be $F_{12}=-F_{21}=B,~ E^\mu=0$. We now consider the response of the current to the perturbations of the electric field $\delta E^{\mu}$. From the anomaly term we can see that the anomalous effect only arises when the magnetic field is parallel to the electric field, so we will focus on the longitudinal electric conductivity in the following, i.e. we consider the perturbation $\delta E^z$.

For the hydrodynamic system, besides the perturbations of the thermodynamic variables
\begin{eqnarray}
\mu(\vec{x},t)&=&\mu+\delta \mu(\vec{x},t)\,,\\
T(\vec{x},t)&=&T+\delta T(\vec{x},t)\,,\\
u^\mu(\vec{x},t)&=&(1,\delta u_i(\vec{x},t))\,, 
\end{eqnarray}
we also need to consider the following perturbations of the external fields: 
$\delta E^z=\delta F^{0z}=-\delta F^{z0}$, $\delta E^x=\delta F^{0x}+B \delta u^y$ and $\delta E^y=\delta F^{0y}-B\delta u^x$ for the use of calculating electric conductivities. In this system, the variables $\mu$, $T$ and $u^{\mu}$ will respond to external perturbations and other thermodynamic variables follow according to the equation of state.

With these perturbations of the hydrodynamic variables, to linear order  the perturbations of the conserved quantities can be determined as follows 
\begin{eqnarray}
\delta T^{00}&=&\delta \epsilon\,,\\
\delta T^{0i}&=&(\epsilon+p)\delta u^i\,, \\
\delta T^{ij}&=&\delta p g^{ij}-\eta\big(\partial_i \delta u_j+\partial_j \delta u_i-\frac{2}{3}g^{ij}\partial_k\delta u_k\big)-\zeta g^{ij}\partial_k\delta u_k\,, \\
\delta J^t&=&\delta \rho+\sigma_B B\delta u_z\,, \\
\delta J^x&=&\rho\delta u_x+\sigma_E\big(\delta F^{0x}+B\delta u^y\big)-\sigma_E T\partial_x\bigg(\delta \frac{\mu}{T}\bigg)\,, \\
\delta J^y&=&\rho\delta u_y+\sigma_E\big(\delta F^{0y}-B\delta u^x\big)-\sigma_E T\partial_y\bigg(\delta \frac{\mu}{T}\bigg)\,, \\
\delta J^z&=&\rho\delta u_z+\sigma_E\delta E_z-\sigma_E T\partial_z\bigg(\delta \frac{\mu}{T}\bigg)+\delta\sigma_BB\,,
\end{eqnarray} 
where $\delta p$ and $\delta \epsilon$ are fully determined by $\delta \mu$ and $\delta T$ and we ignored the chiral vortical effects as they do not affect the result. $\sigma_E$ may also depend on the chemical potential or temperature, which does not need to be the same as the zero density value. When there is no background electric field, the exact expression of $\sigma_E$ does not affect our result. In hydrodynamics the evolution of these perturbations can be determined from the conservation equations.  To take into account the effect of dissipations we introduce the following dissipation terms in the perturbation of the conservation equations of $\delta J_{\mu}$ and $\delta T_{\mu\nu}$:
\begin{eqnarray}\label{conserveperturb}
\partial_{\mu}\delta T^{\mu 0}&=&\delta F^{0z}J_z+\frac{1}{\tau_e}\delta T^{\mu 0}u_{\mu}\,,\nonumber\\
\partial_{\mu}\delta T^{\mu i}&=&\rho\delta F^{0i}+F^{i\lambda}\delta J_{\lambda}+\frac{1}{\tau_m}\delta T^{\mu i}u_{\mu}\,, \\
\partial_{\mu}\delta J^{\mu}&=&c\delta E^\mu B_\mu+\frac{1}{\tau_c}\delta J^\mu u_\mu\nonumber\,,
\end{eqnarray} where $\tau_e$ is the energy relaxation time, $\tau_m$ denotes the momentum relaxation time and $\tau_c$ is the charge relaxation time. In principle, for anisotropic systems $\tau_m$ can be different in different directions and here for simplicity we choose it to be isotropic.
We also emphasize that the relaxation terms act only on the deviations from equilibrium. The equilibrium state can be one with non-vanishing
energy or charge.

Substituting the perturbations into the conservation equations (\ref{conserveperturb}) we get the following equations for the perturbations $\delta \mu$, $\delta T$ and $\delta u^i$
\begin{eqnarray}\label{consereqns}
\big(\partial_t+\frac{1}{\tau_e}\big)\delta\epsilon+\partial_i\big[(\epsilon+p)\delta u_i\big]- \sigma_B B\delta E_z =0\,;&&\\
\big(\partial_t+\frac{1}{\tau_m}\big)\big[(\epsilon+p)\delta u_x\big]+\partial_x\delta p-\eta \big(\partial_j^2\delta u_x+\frac{1}{3}\partial_x\partial_j\delta u_j\big)
-\zeta\partial_x\partial_j\delta u_j
&=&\nonumber\\
\rho\delta F^{0x}+
B\bigg[\rho\delta u_y+\sigma_E\big(\delta F^{0y}-B\delta u_x\big)-\sigma_E T\partial_y\bigg(\delta \frac{\mu}{T}\bigg)\bigg]\,;&&\nonumber\\
\big(\partial_t+\frac{1}{\tau_m}\big)\big[(\epsilon+p)\delta u_y\big]+\partial_y\delta p-\eta \big(\partial_j^2\delta u_y+\frac{1}{3}\partial_y\partial_j\delta u_j\big)
-\zeta\partial_y\partial_j\delta u_j
&=&\nonumber\\
\rho\delta F^{0y}-B\bigg[\rho\delta u_x+\sigma_E\big(\delta F^{0x}+B\delta u_y\big)-\sigma_E T\partial_x\bigg(\delta \frac{\mu}{T}\bigg)\bigg]\,;&&\nonumber\\
\big(\partial_t+\frac{1}{\tau_m}\big)\big[(\epsilon+p)\delta u_z\big]+\partial_z\delta p-\eta \big(\partial_j^2\delta u_z+\frac{1}{3}\partial_z\partial_j\delta u_j\big)
-\zeta\partial_z\partial_j\delta u_j-\rho\delta E_z
&=&0\,;\nonumber\\
\big(\partial_t+\frac{1}{\tau_c}\big)\big[\delta \rho+\sigma_B B\delta u_z\big]+\partial_i(\rho \delta u_i+\sigma_{E}\delta E_i)-\sigma_{E} T\partial_i^2\big(\delta \frac{\mu}{T}\big)
&&\nonumber\\
+\sigma_{E} B(\partial_x\delta u_y-\partial_y\delta u_x)+\partial_z\delta\sigma_B B-cB\delta E_z &=&0\,.\nonumber
\end{eqnarray}

Though $\delta E^{x,y}$ is not equal to $\delta F^{0x,y}$, the responses of the currents to the two quantities are the same. Note that before introducing the anomaly terms, the coefficient in front of $-\partial_i\delta\mu$ is the same as the coefficient of $\delta E_i$ in the equations using the fact that $\delta p=s\delta T+\rho \delta\mu$, i.e. the coefficients in front of $-\partial_i\delta\mu$ and $\delta E_i$ are only 
different in the anomaly related terms of the first and last equations in (\ref{consereqns}). After introducing $\sigma_B$ there is an extra term in front of $\delta E_z$. If we also keep the chiral vortical anomaly terms $\sigma_V$, there will also be extra terms in front of $-\partial_i\delta\mu$ in the equations above.

By Laplace transforming the equations above in the time direction, we get
\begin{eqnarray}
\omega_e\delta\epsilon-i\delta \epsilon^{(0)}+i(\epsilon+p)\partial_i\delta u_i-i\delta E_z \sigma_B B&=&0\,,\nonumber\\
(\epsilon+p)(\omega_m\delta u_x-i\delta u_x^{(0)})+i\partial_x\delta p-i\eta(\partial_j^2\delta u_x+\frac{1}{3}\partial_x\partial_j\delta u_j)-i\zeta\partial_x\partial_j\delta u_j-iB\delta J_y-i\rho\delta F^{0x}&=&0\,,\nonumber\\
(\epsilon+p)(\omega_m\delta u_y-i\delta u_y^{(0)})+i\partial_y\delta p-i\eta(\partial_j^2\delta u_y+\frac{1}{3}\partial_y\partial_j\delta u_j)-i\zeta\partial_y\partial_j\delta u_j+iB\delta J_x-i\rho\delta F^{0y}&=&0\,,\nonumber\\
(\epsilon+p)(\omega_m\delta u_z-i\delta u_z^{(0)})+i\partial_z\delta p-i\eta(\partial_j^2\delta u_z+\frac{1}{3}\partial_z\partial_j\delta u_j)-i\zeta\partial_z\partial_j\delta u_j-i\rho\delta E_z&=&0\,,\nonumber\\
\big(\omega_c\delta\rho-i\delta \rho^{(0)}\big)+
\sigma_B B\big(\omega_c\delta u_z-i\delta u_z^{(0)}\big)+i\partial_i(\rho \delta u_i+\sigma_{E}\delta E_i)-i\sigma_{E} T\partial_i^2\big(\delta \frac{\mu}{T}\big)
&&\nonumber\\
+i\sigma_E B(\partial_x\delta u_y-\partial_y\delta u_x)+i\partial_z\delta\sigma_B B-ic B\delta E_z
&=&0\,,\nonumber
\end{eqnarray}
where $$\omega_e\equiv \omega+\frac{i}{\tau_e}\,,~~~\omega_m\equiv \omega+\frac{i}{\tau_m}\,,~~~\omega_c\equiv \omega+\frac{i}{\tau_c}\,.$$

As $\delta F^{\mu\nu}$ is an external field, we can choose it to be $\delta F^{0\mu}(t,x^i)=\delta F^{0\mu}e^{-i \omega t+i k_i x^i}.$ After a Laplace transformation in the time direction and a Fourier transformation in the spatial direction, we have $\delta E_z=\delta E_z^{(0)}$. Performing a Fourier transform in spatial directions and taking the limit $k\to 0$, we have
\begin{eqnarray}\label{tosolve1}
\omega_e\delta\epsilon-i\delta \epsilon^{(0)}-i\delta E_z \sigma_B B&=&0\,,\\
(\epsilon+p)(\omega_m\delta u_x-i\delta u_x^{(0)})%+i \partial_x \delta p
-i\rho\delta F^{0x} %\nonumber\\  
-iB\big(\rho\delta u_y+\sigma_E(\delta F^{0y}%-\partial_y \delta \mu
- B\delta u_x%+\mu\frac{\partial_y \delta T}{T}
)\big)
&=&0\,,\\
(\epsilon+p)(\omega_m\delta u_y-i\delta u_y^{(0)})%+i\partial_y \delta p
-i\rho\delta F^{0y}%\nonumber\\
+iB\big(\rho\delta u_x+\sigma_E(\delta F^{0x}%-\partial_x \delta \mu
+B\delta u_y
%+\mu\frac{\partial_x \delta T}{T}
\big)&=&0\,,\\
(\epsilon+p)(\omega_m\delta u_z-i\delta u_z^{(0)})-i\rho\delta E_z%+i\partial_z \delta p
&=&0\,,\\\label{tosolven}
\omega_c\delta\rho-i\delta \rho^{(0)}+
\sigma_B B\big(\omega_c\delta u_z-i\delta u_z^{(0)}\big)-ic B\delta E_z&=&0\,.
\end{eqnarray}

Before proceeding we have the dependence of $\delta \epsilon$, $\delta \rho$ and $\delta p$ on $\delta \mu$ and $\delta T$ as 
\begin{eqnarray}
\delta\epsilon&\equiv&e_1\delta\mu+e_2\delta T=\big(\frac{\partial\epsilon}{\partial \mu}\big)\Big{|}_T\delta \mu+\big(\frac{\partial\epsilon}{\partial T}\big)\Big{|}_\mu\delta T\,,\\
\delta\rho&\equiv&f_1\delta\mu+f_2\delta T=\big(\frac{\partial\rho}{\partial \mu}\big)\Big{|}_T\delta \mu+\big(\frac{\partial\rho}{\partial T}\big)\Big{|}_\mu\delta T\,,\\
\delta p&=&\rho\delta \mu+s \delta T\,,
\end{eqnarray} 
where the coefficients $e_1, ~e_2,~f_1$ and $f_2$ are thermodynamic coefficients which depend on the details of different systems. For the longitudinal direction, solving $\delta\mu,\delta \rho,\delta u_z$ in terms of
$\delta\mu^{(0)},\delta \rho^{(0)},\delta u_z^{(0)},\delta E_z^{(0)}$ from (\ref{tosolve1}) to (\ref{tosolven}), we have 
\begin{eqnarray}
\delta u_z&=&\frac{\rho}{\epsilon+p}\frac{i}{\omega_m}\delta E_z^{(0)}+\dots\\\label{deltamu}
\delta \mu&=&\frac{B}{(e_2 f_1-e_1 f_2)}\bigg(-f_2\sigma_B\frac{i}{\omega_e}-
\frac{\rho\sigma_Be_2}{\epsilon+p}\frac{i}{\omega_m}+ce_2\frac{i}{\omega_c}
\bigg)\delta E_z^{(0)}+\dots\\
\delta T&=&\frac{B}{(e_2 f_1-e_1 f_2)}\bigg(f_1\sigma_B\frac{i}{\omega_e}+
\frac{\rho\sigma_Be_1}{\epsilon+p}\frac{i}{\omega_m}-ce_1 \frac{i}{\omega_c}
\bigg)\delta E_z^{(0)}+\dots
\end{eqnarray}
with ``\dots" denoting terms unrelated to $\delta E_z^{(0)}$ which will vanish after we choose the initial values of other perturbations to be zero.

Substituting these evolutions into 
\begin{equation}
\delta J^z=\rho\delta u_z+\sigma_E\delta E_z-\sigma_E T\partial_z\bigg(\delta \frac{\mu}{T}\bigg)+\frac{1}{2}\sigma_V\big(\partial_x \delta u_y-\partial_y\delta u_x\big)+\delta\sigma_BB\,,
\end{equation}
we get (in the $k\to 0$ limit)
\begin{eqnarray}
\delta J^z&=&\rho\delta u_z+\sigma_E\delta E_z+B\bigg(c-\frac{2c\mu\rho+(c\mu^2+c_g T^2)f_1}{2(\epsilon+p)}+\frac{\rho(c\mu^2+c_g T^2)(e_1+\rho)}{2(\epsilon+p)^2}\bigg)\delta \mu \nonumber\\
&&+B\left(-\frac{(c\mu^2+c_g T^2)f_2+2c_g \rho T}{2(\epsilon+p)}+\frac{\rho(c\mu^2+c_g T^2)(e_2+s)}{2(\epsilon+p)^2}\right)\delta T\\
&=&\sigma \delta E_z^{(0)}+\dots
\end{eqnarray}
 with 
\begin{eqnarray}\label{conductivity}
\sigma&=&\sigma_E-\frac{i}{\omega+\frac{i}{\tau_e}}\frac{B^2c\sigma_B}{2(e_2f_1-e_1f_2)}Y_0+\frac{i}{\omega+\frac{i}{\tau_m}}\frac{\rho}{\epsilon+p}\bigg[\rho-
\frac{B^2 c\sigma_B}{2(e_2 f_1-e_1 f_2)} Y_1\bigg]\\&&~~~~~+\frac{i}{\omega+\frac{i}{\tau_c}}\frac{B^2c^2}{2(e_2 f_1-e_1 f_2)}Y_1\nonumber
\end{eqnarray}
where
\begin{eqnarray}
%Y_0&=&2 f_2-\frac{2f_2\mu\rho}{\epsilon+p}+\frac{\mu^2\rho(e_1 f_2-e_2f_1-f_1 s)+f_2\mu^2\rho^2}{(\epsilon+p)^2}\nonumber\\
%&&~~~~+\frac{c_g T \rho}{c(\epsilon+p)}\left[2f_1 -\frac{(e_2f_1- e_1f_2)+(f_1 s-f_2\rho)}{\epsilon+p}T\right],\\
Y_0&=&\frac{1}{\epsilon+p}\bigg[2f_2 Ts -\frac{(e_2f_1- e_1f_2)+(f_1 s-f_2\rho)}{\epsilon+p} \mu^2 \rho\bigg]\nonumber \\
&&+\frac{c_g }{c(\epsilon+p)}\left[2f_1 T \rho -\frac{(e_2f_1- e_1f_2)+(f_1 s-f_2\rho)}{\epsilon+p}T^2\rho \right]\,,\\
%Y_1&=&2e_2+\frac{-(e_2 f_1-e_1f_2)\mu^2-2e_2\mu\rho}{\epsilon+p}
%-\frac{\mu^2\rho(e_1 s-e_2\rho)}{(\epsilon+p)^2}\nonumber\\&&~~~~+\frac{c_g T}{c(\epsilon+p)}\left[2e_1 \rho-(e_2 f_1-e_1f_2) T-\frac{e_1 s-e_2\rho}{\epsilon+p}T\rho\right]\\
Y_1&=&\frac{1}{\epsilon+p}\bigg[ 2 e_2 Ts- (e_2 f_1-e_1f_2) \mu^2 -\frac{e_1 s-e_2\rho}{\epsilon+p}\mu^2\rho\bigg]\nonumber\\~~~~&&+\frac{c_g }{c(\epsilon+p)}\left[2e_1T \rho-(e_2 f_1-e_1f_2) T^2-\frac{e_1 s-e_2\rho}{\epsilon+p}T^2\rho\right]\,.
\end{eqnarray}

Eq. (\ref{conductivity}) is our final result for {\em longitudinal} electric conductivity of a chiral anomalous fluid with background magnetic field. This is a universal hydrodynamic result which applies in the hydrodynamic limit regardless of the microscopic details. When $B=0$ or $c=c_g=0$, the result reduces to 
\begin{equation}
\sigma=\sigma_E+\frac{i}{\omega_m}\frac{\rho^2}{\epsilon+p}\,,
\end{equation}
which is the result for the usual electric conductivity without background magnetic field.  For the electric conductivity in the transverse directions, the effect of magnetic field is similar to that in 2+1 dimensions and the results for $\sigma_{xx}$, $\sigma_{yy}$ as well as $\sigma_{xy}$ in this case are exactly the same as in \cite{nernsteffect}. For future reference, we list the results here:
\begin{equation}\label{sigma-transverse}
\sigma_{xx}=\sigma_{yy}=\sigma_E\frac{\omega_m(\omega_m+i\gamma+i\omega_{cy}^2/\gamma)}{(\omega_m+i\gamma)^2-\omega_{cy}^2}\,,~~~
\sigma_{xy}=-\frac{\rho}{B}\frac{\gamma^2+\omega_{cy}^2-2i\gamma\omega_m}{(\omega_m+i\gamma)^2-\omega_{cy}^2}\,,
\end{equation}
where $\omega_{cy}=\frac{B\rho}{\epsilon+p}$ is the cyclotron frequency and $\gamma=\frac{\sigma_E B^2}{\epsilon+p}$.

There is a lot of information in the longitudinal result (\ref{conductivity}):
\begin{itemize}

\item We can see that the final result is  
related to all the three relaxation times $\tau_e$, $\tau_m$ and $\tau_c$,
i.e. energy relaxation, momentum relaxation and charge relaxation all enter the final result.  This means that all of these three kinds of dissipations are needed to have a finite longitudinal DC conductivity for the chiral anomalous systems. The conductivity instead of the resistivity is a sum of various contributions, which means that here the conductivity satisfies the inverse Matthiesen rule.

\item The energy dissipation and momentum dissipation terms are always associated with the finite charge density.

\item The explicit value of (\ref{conductivity}) depends on the thermal state that the system is in. The thermodynamic quantities also may depend on $B$ so the dependence of this conductivity on $B$ may differ for different systems or in different limits.

\item We can take the $\tau_e\to \infty$, $\tau_m\to \infty$ and $\tau_c\to \infty$ limit to get the result without any relaxation terms. Then we have the following longitudinal conductivity \begin{equation}
\sigma_{zz}=\sigma_E+\frac{i}{w} \bigg[\frac{\rho^2}{\epsilon+P}-\frac{B^2c\sigma_B}{2(e_2f_1-e_1f_2)}Y_0+\frac{B^2c^2Y_1}{2(e_2f_1-e_1f_2)}\bigg(1-\frac{\mu\rho}{\epsilon+P}+\frac{\mu^2\rho^2}{2(\epsilon+P)^2}\bigg)\bigg]\,,\nonumber
\end{equation}which has a pole in the imaginary part at $\omega=0$ and accordingly there will be a $\delta (\omega)$ in the real part of the conductivity.

There are several component parts for this infinite DC conductivity and the different origins of these parts are more easily seen in the relaxed form (\ref{conductivity}). The first is the usual infinite DC conductivity coming from the acceleration of the charge carriers with a charge density $\rho$. As can be seen from the expression of $\delta J^z$, this term is related to $\delta u^z$, which means that it comes from the infinite momentum increase under an external perturbation of the electric field.  The second part is $\frac{i}{\omega}\frac{B^2c^2}{2(e_2 f_1-e_1 f_2)}Y_1$, the third part is $-\frac{i}{\omega}\frac{\rho}{\epsilon+p}
\frac{B^2 c\sigma_B}{2(e_2 f_1-e_1 f_2)} Y_1$ and the fourth term is $-\frac{B^2c\sigma_B}{2(e_2f_1-e_1f_2)}Y_0$. From the calculations we can see that these three terms all come from the response of the chiral magnetic current $\sigma_B B$ to an external electric field and only exist at nonzero values of $c$. They are related to the infinite increase of the chemical potential and the temperature of the system under an external longitudinal electric field.

\item Without the anomaly terms, $\delta \mu$ would not respond to $\delta E^z$ which means that there will be 
no increase of chemical potential or charge density in an anomalous free system. 
The second part can only be dissipated by the momentum relaxation while the third part can only be dissipated by the charge dissipation and the last term can only be dissipated by energy dissipation. This means that the infinite increase of chemical potential or temperature is also related to the increase of momentum or energy under the external electric field.

\item Note that \begin{equation}e_2f_1-e_1f_2=\det
\left(
\begin{array}{ccc}
 \frac{\partial\epsilon}{\partial T} & \frac{\partial\epsilon}{\partial\mu}    \\
 \frac{\partial\rho}{\partial T} & \frac{\partial\rho}{\partial T}  \end{array}
\right),
\end{equation} and it is positive definite. Though the signs of $Y_0$ or $Y_1$ are not easy to tell and may depend on different systems, there is an interesting sign difference in these two parts related to $Y_1$ and though the sign in the momentum relaxation related term looks dependent on the sign of charge, in fact due to the dependence of $\sigma_B$ on the chemical potential, the sign difference does not depend on the sign of the charge.
Also it is easy to check that the absolute value of the second part as defined above is always larger than the absolute value of the third part. Because the sign of $Y_1$ cannot be determined, it may happen that the signs of the two terms in the momentum dissipation related part can either be the same or opposite.

\item Here another interesting point is that there exists a possibility that there might be a state of a certain system at which the divergence term will vanish due to cancellations if the signs and values of the quantities can be fine tuned to appropriate values.

\item Note that even at zero density %frequency 
there can still be an infinite DC conductivity \cite{Jimenez-Alba:2014iia}\footnote{We set $c_g=0$ and we consider the system with $\rho(\mu=0,T)=0$, i.e. $f_2=0$. } 
\begin{equation}\label{zerodensitysigma}
\sigma = \sigma_E+\frac{i}{\omega}\frac{B^2c^2}{(\partial\rho/\partial \mu)|_T}\,,
\end{equation}
and this term can only be dissipated by the charge dissipation. The mechanism of an infinite DC conductivity even at zero frequency has already been explained at the beginning of this section.

 \item We can compare this result with the dissipation terms that were used in literature within the weakly coupled kinetic framework.  
The longitudinal magnetoconductivity for Weyl metal and Weyl semi-metals have been studied in \cite{{Nielsen:1983rb}, {sonspivak}, {kim-boltzman},{gorbar}} using weakly coupled kinetic theories,  Boltzmann equation approach and Kubo formulae. In these calculations, dissipation effects from intra valley and inter valley scatterings have been included so that the final DC conductivity is finite. Comparing their results with the formula obtained in this paper, we can see that the intra valley scattering inside one Weyl cone leads to momentum relaxation while the inter valley scattering which happens between Weyl cones relaxes the charge, momentum and also energy.
We will comment on the energy dissipation in section \ref{sec-con}.

\end{itemize}

 In the next section we will apply the formula for the anomalous magnetoconductivity to a simple holographic model. Then we will check
its validity by computing it directly via the Kubo formula in the same model.

%%%%%%%%%%%%%%%%%%%%%%%%%%%%%%%%%%%%%%%%%%%%%
\subsection{Applying the formula to the holographic probe system}
\label{secapply}
%%%%%%%%%%%%%%%%%%%%%%%%%%%%%%%%%%%%%%%%%%%%%

In this subsection, we apply the formula (\ref{conductivity}) to the simplest holographic system: the Schwarzschild black hole with a nontrivial gauge field in the probe limit, which corresponds to the small density limit as the density of charge carriers is extremely small compared to the density of neutral degrees of freedom. In the probe limit, we are in the high temperature regime\footnote{By probe limit we mean the backreaction of the gauge field is totally unimportant for the gravity background. Thus the high temperature limit means $T \gg \frac{\kappa}{e}\mu$ with $e=1$ in our setup and in this case the contribution of the charged  d.o.f is totally unimportant to the energy momentum tensor in the whole spacetime at leading order. }
as when the temperature gets lower  backreaction onto the geometry becomes more important.  We will first calculate the background of a chiral anomalous system with only one $U(1)$ current and obtain the thermodynamic quantities of this system. Then we can substitute them into the formula (\ref{conductivity}) to get the prediction of the hydrodynamic result to the holographic systems.

The bulk action which corresponds to a chiral anomalous fluid is the AdS Einstein-Maxwell-Chern-Simons
\begin{equation}
{\mathcal S}=\int d^5x\sqrt{-g}\bigg[\frac{1}{2\kappa^2}\Big(R+\frac{12}{L^2}\Big)-\frac{1}{4}F^2+\frac{\alpha}{3}\epsilon^{\mu\nu\rho\sigma\tau}A_\mu F_{\nu\rho} F_{\sigma\tau}\bigg]
\end{equation}
where $\epsilon_{\mu\nu\rho\sigma\tau}=\sqrt{-g}\varepsilon_{\mu\nu\rho\sigma\tau}$ with $\varepsilon_{0123r}=1.$ For simplicity we choose the gravitational anomaly term to be zero.  In the probe limit, which corresponds to a small density system, the background is the AdS Schwarzschild black hole
\begin{equation}
ds^2=r^2\Big(-f(r)dt^2+dx^2+dy^2+dz^2\Big)+\frac{dr^2}{r^2f(r)}
\end{equation}
with $f(r)=1-\frac{r_0^4}{r^4}$ and we set $L=1.$ 
The known thermodynamical quantities are 
\begin{equation}
\epsilon= 3 r_0^4\,,~~s=4\pi r_0^3\,,~~T=\frac{r_0}{\pi}\,.
\end{equation}

 The equation of motion for the gauge field is 
\begin{equation}
\nabla_\nu F^{\nu\mu} +\alpha \epsilon^{\mu\nu\rho\sigma\tau} F_{\nu\rho} F_{\sigma\tau}=0\,.
\end{equation} 
For a background with a magnetic field, we choose $F_{12}=-F_{21}=B$. We consider a nonzero $A_t$ component and an $A_z$ component which couple to each other. This is because with a magnetic field in the $z$ direction and the Chern-Simons term, a current will be excited in the $z$ direction. We will assume $A_{t}=A_t(r)$, $A_{y}=B x$, $A_z=A_z(r)$ and  impose the boundary condition for $A_z$: $A_z(r\to\infty)=\frac{A_z^{(1)}}{r^2}+\cdots$.\footnote{This means that we have set an arbitrary integration constant $A_z^{(0)}$ to zero.}  Note that we will still have a nontrivial solution of $A_t$ here which means that we are not at the strict zero density while almost zero density where the charge density is at $1/N^2$ order compared to neutral degrees of freedom of the system. The equations of motion for these two background gauge fields are
\begin{equation}
A_t''+\frac{3}{r} A_t'+\frac{8B\alpha}{r^3} A_z'=0
\end{equation}
and 
\begin{equation}
A_z''+\frac{3 r^4+r_0^4}{r\big(r^4-r_0^4\big)} A_z'+\frac{8 B\alpha r }{r^4-r_0^4}A_t'=0\,.
\end{equation}
These two equations can be simplified to
\begin{equation}\label{bgeqnataz} 
\big(r^3 A_t'+8B\alpha A_z\big)'=0\,,~~~~~~ r^3(1-\frac{r_0^4}{r^4})A_z'+8B\alpha A_t=0\,.
\end{equation}
Note that in the second equation above, we have used $A_t(r_0)=0.$

We can solve for $A_t$ from these equations analytically and read the corresponding charge density and chemical potential. In the new coordinate $u={r_0^2}/{r^2}$, we have 
\begin{equation}
A_t''-\frac{(8B\tilde{\alpha})^2}{4(1-u^2)} A_t=0\,.
\end{equation}
We have defined $\tilde{\alpha}=\alpha/(\pi^2T^2)$ and the anomalous coefficient is related to the Chern-Simons coupling by $c=8\alpha.$\footnote{We are using {\em covariant current} \cite{Landsteiner:2012kd}, i.e. our dual current is identified as the subleading term of $A_\mu$ near the boundary.} Note that $\alpha$ is dimensionless and $\tilde{\alpha}$ is of the dimension mass$^{-2}$. The analytic solution with the near horizon boundary condition $A_t(u=1)=0$ is
\begin{eqnarray}\label{atbackground}
A_t&=&  \ _2F_1\bigg[-\frac{1+\sqrt{1-(8B\tilde{\alpha})^2}}{4},-\frac{1-\sqrt{1-(8B\tilde{\alpha})^2}}{4},\frac{1}{2},u^2\bigg]
\nonumber\\&&-2u\frac{\Gamma\big[\frac{5-\sqrt{1-(8B\tilde{\alpha})^2}}{4}\big]\Gamma\big[\frac{5+\sqrt{1-(8B\tilde{\alpha})^2}}{4}\big]}{\Gamma\big[\frac{3-\sqrt{1-(8B\tilde{\alpha})^2}}{4}\big]\Gamma\big[\frac{3+\sqrt{1-(8B\tilde{\alpha})^2}}{4}\big]}\
_2F_1\bigg[\frac{1-\sqrt{1-(8B\tilde{\alpha})^2}}{4},\frac{1+\sqrt{1-(8B\tilde{\alpha})^2}}{4},\frac{3}{2},u^2\bigg]\nonumber\,.
\end{eqnarray} 
Near the boundary $u=0$ this solution should behave as $A_t=\mu -\frac{\rho}{2r_0^2}u+\dots$, thus we can expand the solution (\ref{atbackground}) at the boundary $u=0$ and obtain the value of the dual charge density as
\begin{equation}\label{rhomuinsch}
\rho=4\mu r_0^2 \frac{\Gamma\big[\frac{5-\sqrt{1-(8B\tilde{\alpha})^2}}{4}\big]\Gamma\big[\frac{5+\sqrt{1-(8B\tilde{\alpha})^2}}{4}\big]}{\Gamma\big[\frac{3-\sqrt{1-(8B\tilde{\alpha})^2}}{4}\big]\Gamma\big[\frac{3+\sqrt{1-(8B\tilde{\alpha})^2}}{4}\big]}\,.
\end{equation}

Up to now we have at hand all the thermodynamic quantities that are needed in (\ref{conductivity}) for this probe case without any dissipation effects, except for $\sigma_E$, which we will give in the next section as it involves perturbative analysis. Substituting (\ref{rhomuinsch}) into (\ref{zerodensitysigma}) we get
\begin{eqnarray}\label{probeprediction}
\sigma_{zz}&=&\sigma_E+\frac{i}{\omega}\frac{B^2c^2}{(\partial\rho/\partial \mu)|_T}\nonumber\\\label{confromhydro-hol}
&=&\sigma_E+\frac{i}{\omega}\frac{B^2c^2}{4\pi^2 T^2}\frac{\Gamma\big[\frac{3-\sqrt{1-(8B\tilde{\alpha})^2}}{4}\big]\Gamma\big[\frac{3+\sqrt{1-(8B\tilde{\alpha})^2}}{4}\big]}{\Gamma\big[\frac{5-\sqrt{1-(8B\tilde{\alpha})^2}}{4}\big]\Gamma\big[\frac{5+\sqrt{1-(8B\tilde{\alpha})^2}}{4}\big]}\,.
\end{eqnarray}
Eq. (\ref{probeprediction}) is the explicit result for the formula (\ref{conductivity}) applied in the holographic system in the zero density limit with the number density of the charge carriers $n_{c}\ll n_\text{neutral}$.

The $B$ dependence of this result involves a lot of $\Gamma$-functions. Note that the hydrodynamic formula Eq. (\ref{conductivity}) is only valid for $B\ll T^2$, so Eq. (\ref{probeprediction}) is also in the regime $B\ll T^2$. We can see that in the expression (\ref{probeprediction}) the $B$ dependence is always encoded in the combination of $B\alpha/T^2$, and as $\alpha$ typically is of the order of $N_c$ \cite{Gynther:2010ed} for fermions in the fundamental representation,\footnote{In holography, we can add fermions in the fundamental representation by adding $N_f$ spacetime filling probe branes in the background of $N_c$ D$_3$ branes and as $N_f\ll N_c$, the backreaction can be ignored.} i.e. $\alpha\gg 1$ in the large $N_c$ limit, we have another scale here: $T^2/\alpha$, which is much smaller compared to $T^2$.  In the hydrodynamic regime, we assume both $B\ll T^2$ and $\alpha B\ll T^2$. In this limit, i.e. for $\alpha B/T^2 \ll 1,$ we have
\begin{equation}\label{eq:rhoexpansion}
\rho=2\mu r_0^2\bigg(1+\mathcal{O}\big(\frac{\alpha^2 B^2}{T^4}\big)\bigg)\,,
\end{equation}
and 
\begin{equation}\label{smallB}
\sigma=\sigma_E+\frac{i}{\omega}\bigg(\frac{B^2c^2}{2\pi^2 T^2}+\dots\bigg)\,.\end{equation} 
Note that $\sigma_E$ may also depend on $B$ in a nontrivial way, so we will only get the full dependence of the magnetoconductivity on $B$ in the next section.

%%%%%%%%%%%%%%%%%%%%%%%%%%%%%%%%%%%%%%%%%%%%%%%%
\section{Holographic magnetoconductivity for chiral anomalous fluid in the probe limit 
via Kubo formula}
\label{sec3}
%%%%%%%%%%%%%%%%%%%%%%%%%%%%%%%%%%%%%%%%%%%%%%%%

In this section, we will focus on the direct holographic calculation of the infinite DC magnetoconductivity for chiral anomalous fluid in the probe limit.\footnote{The holographic magnetoconductivity for 2+1D strange metals has also been studied in e.g. \cite{{nernsteffect},{elias}}.} We will calculate the conductivity in this holographic background and check that it is the same as the hydrodynamic prediction of the last section. In particular we will consider small perturbations on top of the background and compute the longitudinal conductivity directly from the perturbations. At the end we will show that the result of this subsection exactly matches the result in the last subsection from the application of the formula (\ref{conductivity}) to the same holographic system. We will also get the exact expression for $\sigma_{Ez}$ which is crucial for the full behavior of the longitudinal magnetoconductivity. Note that in the holographic model we can go beyond the hydrodynamic approximation and therefore allow for anisotropic quantum critical conductivities, singling out $\sigma_{Ez}$ as the longitudinal one.

To compute the longitudinal conductivity, we turn on the fluctuations $\delta A_t(r) e^{-i\omega t}$, $\delta A_z(r) e^{-i\omega t}$. This is because $\delta A_z(r)$ will also source $\delta A_t(r)$ due to the anomaly related terms. This is consistent with the calculations in (\ref{deltamu}) that $\delta \mu$ can be sourced by $\delta E_z$.

The equations of motion for the perturbations are 
\begin{eqnarray}
\frac{8\alpha B}{r^3}\delta A_z+\delta A_t'&=&0\,,
\\
\delta A_z''+\frac{(3r^4+r_0^4)}{r(r^4-r_0^4)}\delta A_z'
+\frac{r^4\omega^2}{(r^4-r_0^4)^2}\delta A_z+\frac{8\alpha Br}{r^4-r_0^4}\delta A_t'&=&0\,.
\end{eqnarray} We can eliminate $A_t$ in the equations and we have
\begin{equation}
 \delta A_z''+\frac{3+\frac{r_0^4}{r^4}}{r\big(1-\frac{r_0^4}{r^4}\big)} \delta A_z'+\bigg(-\frac{64\alpha^2B^2}{r^6\big(1-\frac{r_0^4}{r^4}\big)}+\frac{\omega^2}{r^4\big(1-\frac{r_0^4}{r^4}\big)^2}\bigg) \delta A_z=0\,.
\end{equation}Note that the equations for $\delta A_t, \delta A_z$ do not depend on the background $A_t$ or $A_z$.

In the new coordinate $u=\frac{r_0^2}{r^2}$, we have 
\begin{equation}\label{holconeom}
\delta A_z''-\frac{2u}{1-u^2} \delta A_z'+\bigg(
-\frac{16\alpha^2B^2}{r_0^4(1-u^2)}+\frac{\omega^2}{4r_0^2u(1-u^2)^2}\bigg)\delta A_z=0\,.
\end{equation}We can first solve it in the near horizon region with ingoing boundary conditions and then match it to a far region solution. In the near horizon region $1-u\ll 1$, the equation (\ref{holconeom}) becomes
\begin{equation}
\delta A_z^{(n)''}-\frac{1}{1-u} \delta A_z^{(n)'}+\bigg(
-\frac{8\alpha^2B^2}{r_0^4(1-u)}+\frac{\omega^2}{16r_0^2(1-u)^2}\bigg)\delta A_z^{(n)}=0\,.
\end{equation}

The solution to this equation is the modified Bessel function of the first kind
\begin{equation}
\delta {A_z^{(n)}}=c_1(-1)^{-\frac{i\omega}{4r_0}}
I\Big[-\frac{i\omega}{2r_0},\frac{4\sqrt{2}\alpha B(1-u)^{1/2}}{r_0^2}\Big]
+c_2 (-1)^{\frac{i\omega}{4r_0}}
I\Big[\frac{i\omega}{2r_0},\frac{4\sqrt{2}\alpha B(1-u)^{1/2}}{r_0^2}\Big]\,,
\end{equation}
with two integration constants $c_1$ and $c_2$. Here we impose the infalling boundary condition for $u\to 1$ which corresponds to $c_2=0$.

In the far region $1-u\gg \omega/r_0$, the equation becomes
\begin{equation}
\delta {A_z^{(f)}}''-\frac{2u}{1-u^2} \delta {A_z^{(f)}}'+\bigg(
-\frac{16\alpha^2B^2}{r_0^4(1-u^2)}\bigg)\delta {A_z^{(f)}}=0\,.
\end{equation}
The corresponding solution is the Legendre functions of first and second kind
\begin{equation}\label{farsch}
\delta {A_z^{(f)}}=c_3 P\Big[-\frac{1}{2}+\frac{1}{2}\sqrt{1-64\frac{\alpha^2 B^2}{r_0^4}},u\Big]+c_4 Q\Big[-\frac{1}{2}+\frac{1}{2}\sqrt{1-64\frac{\alpha^2 B^2}{r_0^4}},u\Big]\,.
\end{equation}

To determine the integration constants, we need to match this solution to the near horizon solution. In the matching region $\omega/r_0 \ll 1-u\ll 1$ the near horizon solution becomes
\begin{equation}\label{nearsol-maching}
\delta {A_z^{(n)}}=c_1\bigg[\big(1+\mathcal{O}(\omega)\big)-\frac{i\omega}{4r_0}\log(1-u)\big(1+\mathcal{O}(\omega)\big)+\mathcal{O}(1-u)\bigg]\,,
\end{equation}
where the ratio of the coefficients of the two linearly independent solutions $1$ and $\log(1-u)$ of the matching region is $-\frac{i\omega}{4r_0}$, which is only accurate up to leading order in $\omega$ and higher order corrections in $\omega$ require higher order expansions in the equations of motion of the near region, which we do not consider here.

  The far region solution becomes
\begin{equation}\label{farsol-matching}
\delta A_z^{(f)}=c_3+c_4\bigg[-\frac{1}{2}\log (1-u)+\frac{1}{2}C\bigg]+\mathcal{O}(1-u)
\end{equation}in the matching region with
\begin{eqnarray}
C&=&-\log 2-\cos^2\Big(\frac{\pi(1+\sqrt{1-a_0})}{4}\Big)\bigg(H[-\frac{3}{4}-\frac{1}{4}\sqrt{1-a_0}]+H[-\frac{3}{4}+\frac{1}{4}\sqrt{1-a_0}]\bigg)\nonumber\\
&&
-\sin^2\Big(\frac{\pi(1+\sqrt{1-a_0})}{4}\Big)\bigg(H[-\frac{1}{4}-\frac{1}{4}\sqrt{1-a_0}]+H[-\frac{1}{4}+\frac{1}{4}\sqrt{1-a_0}]\bigg)
\end{eqnarray}
where $a_0\equiv 64 \alpha^2 B^2/r_0^4 $ and $H$ is the harmonic number. 

Matching these two solutions (\ref{nearsol-maching}) and (\ref{farsol-matching}), we get
\begin{equation}\label{c3c4}
c_3=c_1\big(1+\mathcal{O}(\omega)\big)\,,~~~c_4=\frac{i\omega}{2r_0}c_1\big(1+\mathcal{O} (\omega)\big)\,.
\end{equation}

Then we get the far region solution which corresponds to the infalling one by substituting these coefficients into the far region solution and the ratio of the two linearly independent solutions is $c_4/c_3=i\omega/(2r_0)$ at leading order in $\omega$. At the boundary $u\to 0$, the two linearly independent far region solutions (\ref{farsch}) can be expanded to give
\begin{equation}\label{LPE}
P\Big[-\frac{1}{2}+\frac{1}{2}\sqrt{1-64\frac{\alpha^2 B^2}{r_0^4}},u\Big]= p_1+p_2 u+\mathcal{O}(u^2)\,,
\end{equation}
where\footnote{Note that we have defined $\tilde{\alpha}=\alpha/r_0^2$.}
 \begin{equation}p_1=\frac{\sqrt{\pi}}{\Gamma\big[\frac{3-\sqrt{1-(8B\tilde{\alpha})^2}}{4}\big]\Gamma\big[\frac{3+\sqrt{1-(8B\tilde{\alpha})^2}}{4}\big]}\,, ~~p_2=-\frac{8\sqrt{\pi}\tilde{\alpha}^2 B^2}{\Gamma\big[\frac{3-\sqrt{1-(8B\tilde{\alpha})^2}}{4}\big]\Gamma\big[\frac{3+\sqrt{1-(8B\tilde{\alpha})^2}}{4}\big]} \nonumber\end{equation}
 and
 \begin{equation}\label{LQE}
Q\Big[-\frac{1}{2}+\frac{1}{2}\sqrt{1-64\frac{\alpha^2 B^2}{r_0^4}},u\Big]= q_1+q_2 u+\mathcal{O}(u^2)\,,
 \end{equation}
where
 \begin{equation}q_1=\frac{\sqrt{\pi}\cos\big[\frac{\pi(1+\sqrt{1-(8B\tilde{\alpha})}}{4}\big]\Gamma\big[\frac{1+\sqrt{1-(8B\tilde{\alpha})^2}}{4}\big]}{2\Gamma\big[\frac{3+\sqrt{1-(8B\tilde{\alpha})^2}}{4}\big]}\,, ~~q_2= \frac{\sqrt{\pi}\sin\big[\frac{\pi(1+\sqrt{1-(8B\tilde{\alpha})}}{4}\big]\Gamma\big[\frac{3+\sqrt{1-(8B\tilde{\alpha})^2}}{4}\big]}{\Gamma\big[\frac{1+\sqrt{1-(8B\tilde{\alpha})^2}}{4}\big]}\,.\nonumber\end{equation}
Thus at the boundary, the solution which satisfies the near horizon ingoing boundary condition behaves as
\begin{equation}\label{azbnd}
\delta A_z= a+bu+\cdots= c_3 p_1+c_4 q_1+u(c_3 p_2+c_4 q_2)+\cdots\,,
\end{equation}where $c_3$ and $c_4$ have already been fixed from (\ref{c3c4}) by the ingoing boundary condition at the horizon.

At the boundary, $\omega^2$ term will introduce $\omega^2$ correction in $b$ and also a $u\ln u$ term in (\ref{azbnd}). After substituting the divergence term, 
the definition of conductivity\footnote{The definition of the current depends on the different formalism  
that we are in. For the {\em consistent current}, i.e. $ 
J^{\mu}=\sqrt{-g}F^{\mu r}+\frac{4}{3}\alpha\epsilon^{\mu\nu\rho\lambda}A_\nu F_{\rho\lambda},$ which follows from the dictionary due to the bulk Chern-Simons term %will change the definition of the current at the boundary to 
\cite{Gynther:2010ed}.  %but 
Here we stay in the covariant definition so the second term is not included. An extra $i\omega /2$ term arises because of removing the logarithmic term in the asymptotic series of $\delta A_z$ by adding the counterterm \cite{Horowitz:2008bn}.} is
\begin{equation}
\sigma=\frac{2r_0^2 b}{i\omega a}+\frac{i \omega}{2}\,,
\end{equation}
 and it is now 
 \begin{equation}\label{confin} \sigma=\frac{2r_0^2 (c_3 p_2+c_4 q_2)}{i\omega (c_3 p_1+c_4 q_1)}+\frac{i\omega}{2}\,. 
 \end{equation}
 
 In general $p_1$, $p_2$, $q_1$ and $q_2$ are order $1$ quantities compared to $\omega$, so we can expand (\ref{confin}) in $c_4/c_3\sim\omega$ as
\begin{equation}
\sigma=\frac{2 r_0^2}{i\omega}\bigg[ \frac{p_2}{p_1}+\frac{c_4}{c_3} \frac{q_2p_1-q_1p_2}{p_1^2}+\mathcal{O}\bigg(\frac{c_4^2}{c_3^2}\bigg) \bigg]\,.
\end{equation}
Substituting (\ref{LPE}) and (\ref{LQE}) into the formula above and we get 
\begin{equation}\label{confromkubo}
\sigma=\bigg[\frac{8\pi\alpha^2 B^2}{r_0^3}\sec\left(\frac{\pi}{2}\sqrt{1-(8B\tilde{\alpha})^2}\right)+\frac{i}{\omega}\frac{16 B^2
\alpha^2}{\pi^2 T^2}\bigg]\frac{\Gamma\big[\frac{3-\sqrt{1-(8B\tilde{\alpha})^2}}{4}\big]\Gamma\big[\frac{3+\sqrt{1-(8B\tilde{\alpha})^2}}{4}\big]}{\Gamma\big[\frac{5-\sqrt{1-(8B\tilde{\alpha})^2}}{4}\big]\Gamma\big[\frac{5+\sqrt{1-(8B\tilde{\alpha})^2}}{4}\big]}
\end{equation}
at leading orders in $\omega.$ Note that as our $c_4/c_3$ is correct at leading order in $\omega$, the result for $\sigma$ will be correct in order $1/\omega$ and $\mathcal{O}(1)$, while at order $\mathcal{\omega}$ there should be corrections in $\sigma$.

The residue of the pole at $\omega=0$ is the same as in (\ref{probeprediction}) and the finite part defines $\sigma_{Ez}$. Note that in the probe limit
there is no dependence on the chemical potential $\mu$.
It is interesting to note that the $i/\omega$ term matches exactly the result from our hydrodynamic linear response computations (\ref{confromhydro-hol}) although  this formula (\ref{confromhydro-hol})  holds only in the hydrodynamic regime\footnote{For simplicity we assume from now on $c,\alpha,B > 0$.} $B\ll T^2$ as well as $\alpha B\ll T^2$. In the holographic case the result holds in a more general regime of $B$, and we can have two interesting regimes in the range $B\ll T^2$: one is the small $B$ limit $B\ll T^2/\alpha$ and the other is an intermediate regime $T^2/\alpha\ll B\ll T^2$. The behavior of the conductivity is very different in these two regimes and
we can extract the $B$ dependence for the small and intermediate $B$ limit analytically by expanding (\ref{confromkubo}) in terms of $\alpha B/T^2$. Note that $\alpha=c/8$ and $\tilde{\alpha}=\alpha/r_0^2$. For small $\alpha B/T^2\ll 1$, we have 
\begin{equation}\label{smallBsigmaE} 
\sigma=\sigma_{Ez}+\frac{i}{\omega}\frac{c^2 B^2}{2\pi^2 T^2}+\mathcal{O}(\alpha^4 B^4)\,,
~~~\sigma_{Ez}=\pi T\left(1-\frac{c^2 B^2 \log 2}{2\pi^4 T^4}+\mathcal{O}(\alpha^4 B^4) \right)\,.
\end{equation}
The leading term in $\sigma_{Ez}$ agrees with previous results without background magnetic field (e.g. \cite{Kovtun:2008kx}).  
Note also that in the hydrodynamic limit the dependence on the magnetic field of $\sigma_{Ez}$ is subleading compared to the frequency dependent term in $\sigma$. Only for frequencies of the order $\omega \sim T$ both terms would be comparable, these are however outside the validity of the hydrodynamic approximation. 

When $\alpha B/T^2\gg 1$, we have 
\begin{equation}\label{largeBsigmaE}
\sigma=\sigma_{Ez}+\frac{i}{\omega} cB+\mathcal{O}\left(\frac{1}{cB}\right),~~~
\sigma_{Ez}=T e^{-\frac{c B}{2\pi T^2}} \left( \frac{cB}{T^2}+\mathcal{O}\left(\frac{1}{cB}\right)\right)\,.
\end{equation} 
Here $\sigma_{Ez}$ also gets nontrivial corrections with $B$.  As pointed out in \cite{{Karch:2007pd},{Donos:2014cya}}, because $\sigma_{Ez}$ exists even at zero density, we can interpret it as coming from the vacuum pair production.

\begin{figure}[h!]
\begin{center}
\begin{tabular}{cc}
\includegraphics[width=0.7\textwidth]{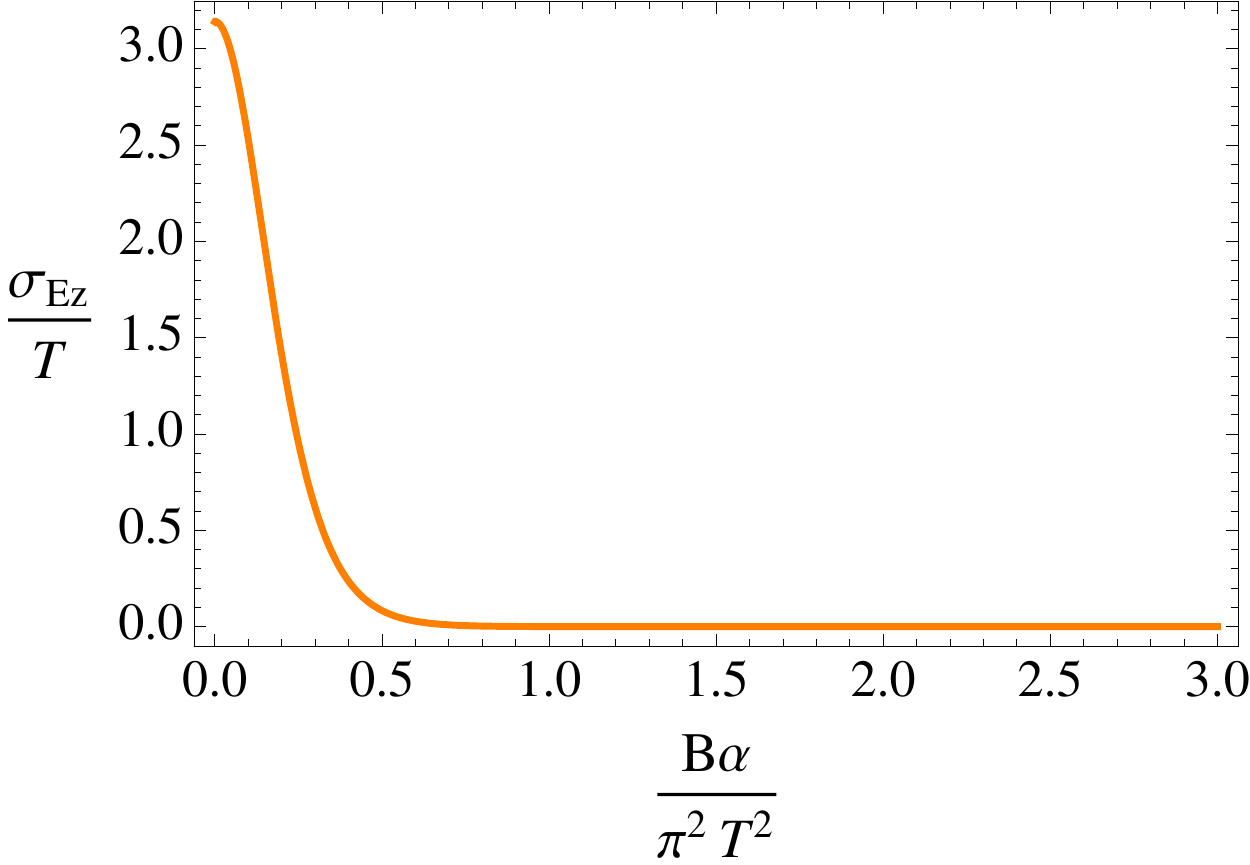}
\end{tabular}
\end{center}
\caption{\small Anomaly related magneto-quenching of the quantum critical conductivity $\sigma_{Ez}$ from (\ref{confromkubo}) as a function of $\alpha B/\pi^2T^2$.}
\label{re-con}
\end{figure}

\begin{figure}[h!]
\begin{center}
\begin{tabular}{cc}
\includegraphics[width=0.7\textwidth]{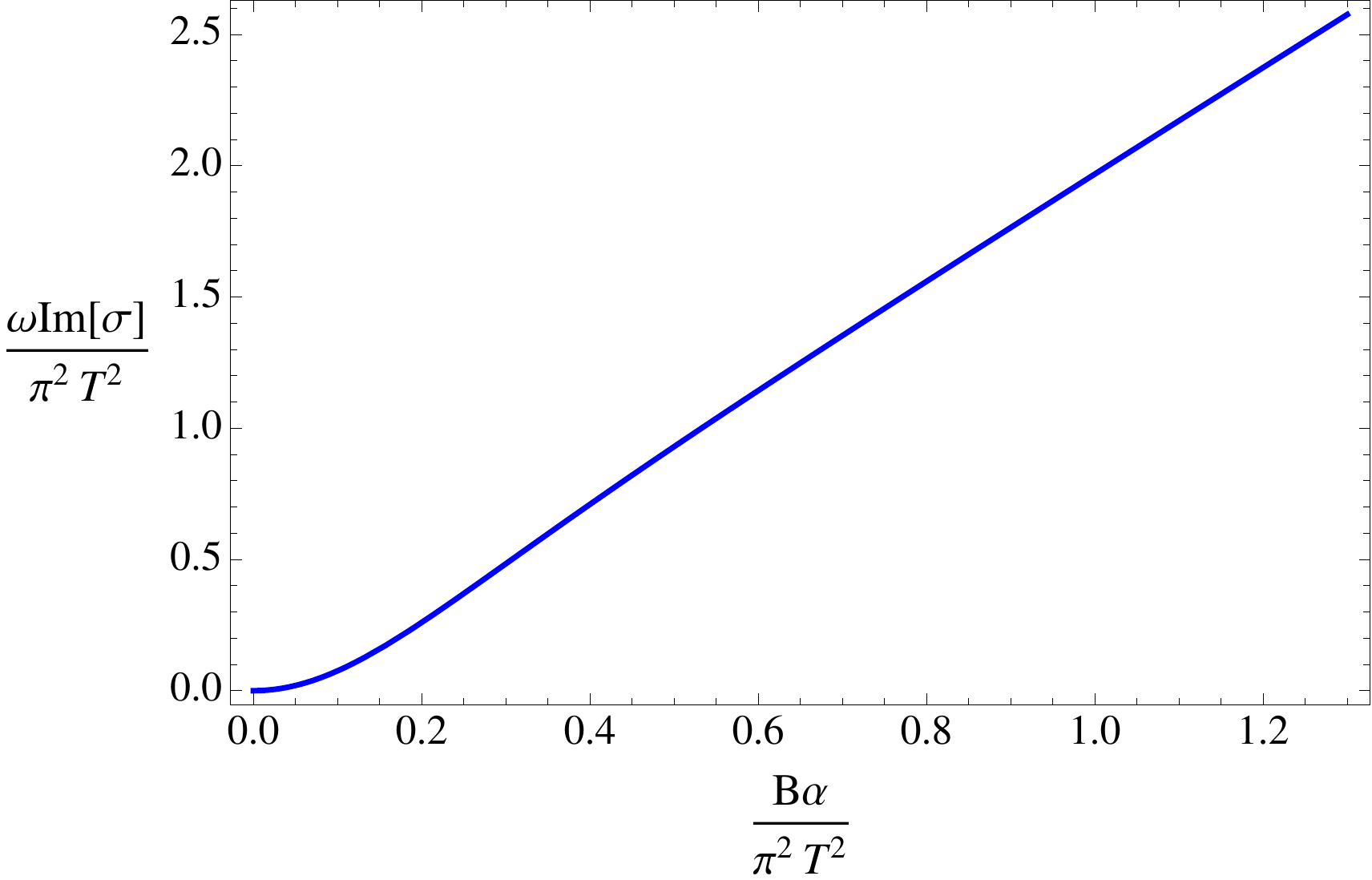}
\end{tabular}
\end{center}
\caption{\small The plot of the dependence of the rescaled coefficient in front of $i/\omega$ in the imaginary part of the conductivity (\ref{probeprediction}) as a function of $\alpha B/\pi^2T^2$. At small $\alpha B/T^2$ it is quadratic in $B$ and at large $\alpha B/T^2$ (intermediate regime) it is linear in $B$.}
\label{b-depend}
\end{figure}

As we already mentioned in the previous section, the expression of $\sigma_{Ez}$ is important for the whole dependence of the longitudinal magnetoconductivity on the magnetic field $B$.  Figure \ref{re-con} shows the behavior of $\sigma_{Ez}$ as a function of $B$.
Notice that the anomaly and the magnetic field quench the quantum critical conductivity for $\alpha B >T^2$.
In figure \ref{b-depend} we plot the dependence of the residue of the DC conductivity at $\omega=0$ on $\alpha B/T^2$. We can see from the picture that at small $B$ it is quadratic in $B$ and at large $\alpha B/T^2$, i.e. the intermediate regime, it is linear in $B$.

%%%%%%%%%%%%%%%%%%%%%%%%%%%%%%%
%%%%%%%%%%%%%%%%%%%%%%%%%%%%%%%%%%%%

%%%%%%%%%%%%%%%%%%%%%%%%%%%%%%%
%%%%%%%%%%%%%%%%%%%%%%%%%%%%%%%%%%%%

We can introduce a charge relaxation time $\tau_c$ in (\ref{confromkubo}) by replacing $\omega$
with $\omega+\frac{i}{\tau_c}.$\footnote{Note that this replacing is not necessarily related to hydrodynamics. Here we use this replacement as a phenomenological attempt to parametric the finite Drude peak.}
%(\ref{smallBsigmaE}) and (\ref{largeBsigmaE}). 
Then the longitudinal DC magnetoconductivity becomes
\begin{equation}
\label{smallBwithtauc} 
\sigma=\sigma_{Ez}+\tau_c\frac{c^2 B^2}{2\pi^2 T^2}+\mathcal{O}(\alpha^4 B^4)\,,~~~\sigma_{Ez}=\pi T-\frac{c^2 B^2 \log 2}{2\pi^3 T^3}+\mathcal{O}(\alpha^4 B^4)
\end{equation}
in the small $B$ limit and 
\begin{equation}\label{largeBwithtauc}
\sigma=\sigma_{Ez}+\tau_c cB+\mathcal{O}\left(\frac{1}{cB}\right)\,,~~~
\sigma_{Ez}=e^{-\frac{c B}{2\pi T^2}}\left( \frac{cB}{T}+\mathcal{O}\left(\frac{1}{cB}\right)\right)\end{equation}
in the intermediate $B$ limit.  The relaxation time $\tau_c$ can depend on the temperature in a nontrivial way. For a fixed nonzero temperature, we can see from (\ref{smallBwithtauc}) that at small $B$, the DC conductivity is dominated by the quantum critical conductivity $\sigma_{Ez}$ while at intermediate $B$ $\sigma_{Ez}\to 0$ and the DC conductivity is dominate by the $\tau_c$ term.
More precisely at small $B$ the magneto-quenching effect in the quantum critical conductivity dominates if the relaxation time obeys
\begin{equation}
 \pi T \tau_c < \log(2)\,.
\end{equation}

\begin{figure}[h]
\begin{center}
\begin{tabular}{cc}
\includegraphics[width=0.7\textwidth]{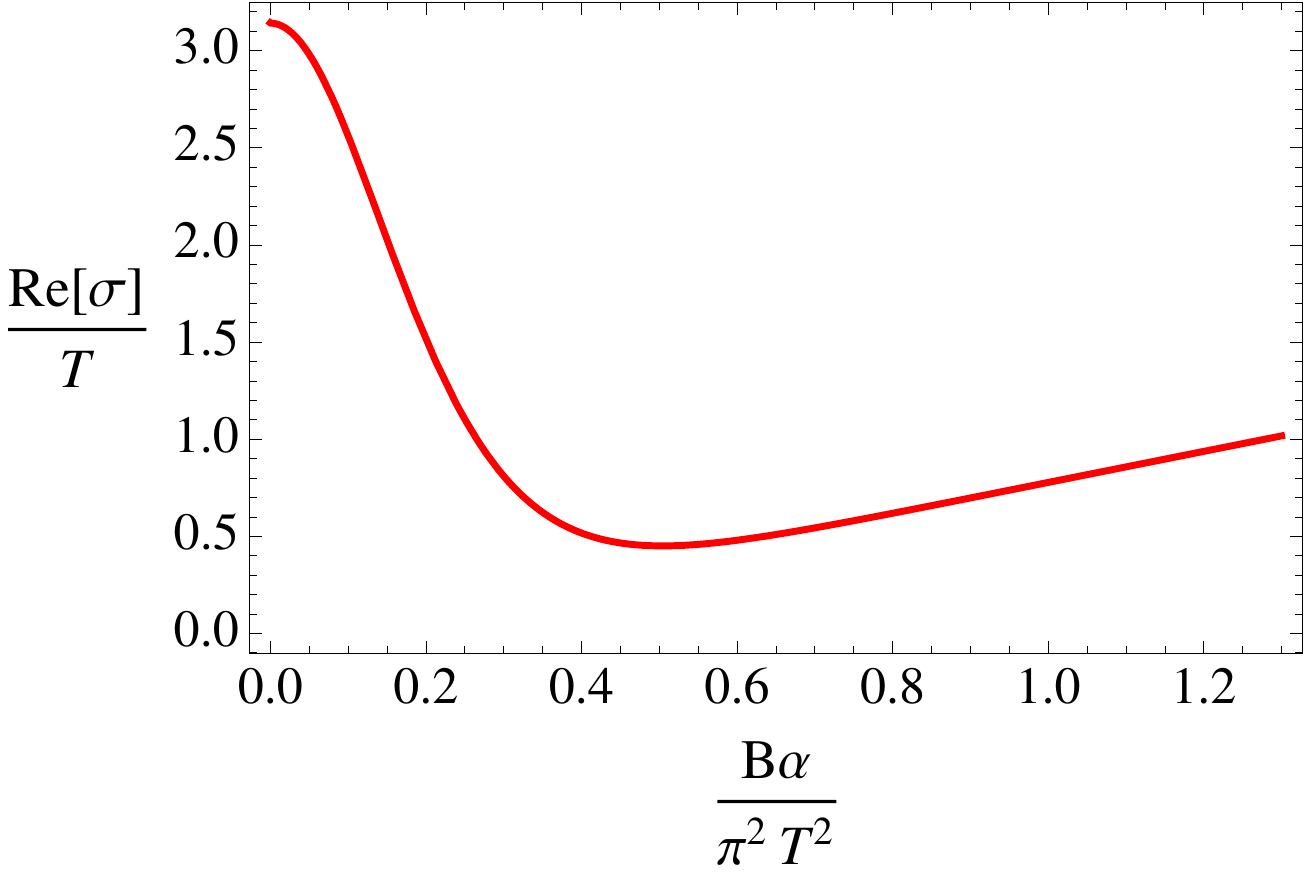}
\end{tabular}
\end{center}
\caption{\small The longitudinal DC magnetoconductivity as a function of the background magnetic field $B$ at a fixed temperature. We have assumed a
charge relaxation time of $T \tau_c = 0.01$. At small $B$ the decreasing of the conductivity with $B$ is caused by the effect of the chiral anomaly in the quantum critical conductivity. In the intermediate regime of $B$ $\sigma_{Ez}\to 0$ and the negative magnetoresistivity is caused by the generation of charge density due to the chiral anomaly effect in the second term of (\ref{largeBwithtauc}). Qualitatively this is the same behavior as seen
in experiments \cite{weyl-exp}.}
\label{negativemr}
\end{figure}

At intermediate $B$ regime, $\sigma\sim B$ which grows linearly in $B$ and this is exactly the negative magnetoresistivity, or equivalently positive magnetoconductivity. At small $B$ for small value of $\tau_c$ $\sigma$ will first decrease with the increase of $B$ and then connect to the negative magnetoresistivity behavior in an intermediate regime of $B$. Fig. \ref{negativemr} shows %is a figure showing
the dependence of $\sigma$ on $B$ at a fixed nonzero temperature where we assumed an appropriate value of $\tau_c$ ($T\tau_c = 0.01$) which behaves as $1/T$ and does not depend on $B$. We can see from the picture that this shows qualitatively the same features as seen in the experimental result of Fig. 3 in \cite{weyl-exp}, namely a negative magnetoresistivity at intermediate regime of $B$ and a decrease of magnetoconductivity as a function of $B$ at small $B$.  The holographic result does however not show a cusp-like behavior near $B=0$ as the derivative of the conductivity at $B=0$ is $0$.

Now we can compare this strongly coupled holographic result with the weak-coupling kinetic result for a Weyl metal in \cite{Nielsen:1983rb,{sonspivak}}. In the limit $\mu,~T\ll \sqrt{B}$, they got the linear in $B$ behavior in the DC conductivity. In the hydrodynamics calculations of Sec. \ref{sec2} we have to stay in the regime $B\ll T^2$. In the holographic calculations we can go to the limit $B\gg T^2, \mu^2$ and the result is the same as found in \cite{Nielsen:1983rb} and \cite{sonspivak} for large $B$. In \cite{sonspivak}, the authors considered the limit $T,~\sqrt{B}\ll \mu$, and they got a $B^2/\mu^2$ behavior, which is different from our small $B$ %case $B^2/T^2$
behavior here (\ref{smallBwithtauc}) as in the probe limit, we cannot go to the zero temperature limit. It would be interesting to also work in the $T\ll\mu$ limit holographically to check the $B$ dependence of the DC conductivity in that limit by considering backreaction of the gauge field, which also will introduce a nontrivial term related to the charge density as can be seen in (\ref{conductivity}).  In \cite{sonspivak}, the limit $\mu,~\sqrt{B}\ll T$ was also considered and the result is the same as our holographic result in the small $B$ limit $\alpha B\ll T^2$.

%%%%%%%%%%%%%%%%%%%%%%%%%%%%%%%
%%%%%%%%%%%%%%%%%%%%%%%%%%%%%%%%%%%

In experiments,  the negative magnetoresistivity behavior caused by chiral anomaly matches well with the weakly coupled field theoretical approach, however, at small $B$ the experimental data does not fit well with theoretical predictions.  In \cite{weyl-exp} the authors proposed to also add the quantum effect of the weak anti-localization, which can fit the experimental results well but is not caused by the chiral anomaly. Here, our holographic results give another possible explanation of the small $B$ behavior,  which comes from the quantum critical conductivity being affected by the chiral anomalous term and does not depend on the charge density. However, there is no cusp-like behavior at $B=0$ as those found in weak anti-localization effects.

Let us now have a look at the behavior of the transverse magnetoconductivity. 
 For the holographic system in the probe limit $\sigma_E$ in the $x,~y$ directions is not affected by the chiral anomalous term, i.e. $\sigma_E=\pi T$
(see appendix).  Thus there is no negative magnetoresistivity as for the longitudinal 
direction.
We note that going beyond the probe limit also the transverse conductivities might pick up some non-trivial $B$ (and $\mu$) dependence.

Let us now briefly discuss the Hall effect.\footnote{Note that if one choses the integration constant for $A_z^{(0)}\neq 0$
the consistent current $J^\mu = \sqrt{-g}F^{\mu r}+\frac{4}{3}\alpha\epsilon^{\mu\nu\rho\lambda}A_\nu F_{\rho\lambda}$ would also show an anomalous (i.e. $B$-field independent) Hall effect proportional to $A_z^{(0)}$. For the Hall effect in the Saki-Sugimoto model see also \cite{Lifschytz:2009si}.}
In the holographic probe limit and on the absence of the Chern-Simons term, though $\rho $ and $B$ should be of the same order, we still cannot see the normal Hall conductivity of (\ref{sigma-transverse}) because backreaction on the gravity need to be considered to couple the two modes of $\delta A_x$ and $\delta A_y$. However with Chern-Simons term we do have a coupling between $\delta A_x$ and $\delta A_y$ due to the chiral anomalous term at order $c^2B$, meanwhile, this is still consistent with the result of (\ref{sigma-transverse}) because this encodes the correction of the chiral anomalous term to the charge density $\rho$ at order $c^2B^2$ resulting in an order $c^2B$ correction in the Hall conductivity. We find
\begin{equation}\label{eq:Hallcond}
 \sigma_{xy} = -\frac{\rho-\rho_h}{B} 
\end{equation}
where $\rho_h$ is the charge density carried by the horizon. 
Note that the Hall conductivity is higher order in $\alpha B/T^2$  and therefore can not be seen in the hydrodynamic theory
of section \ref{sec2}.
More details on the calculation can be found in the appendix. We take this dynamically induced Hall effect as a signature of the the so-called chiral shift shown
to exists at weak coupling in \cite{Gorbar:2009bm}. More precisely a magnetic field induces an interaction driven relative displacement of the longitudinal momenta
of the dispersion relations of fermions of opposite chirality.
In \cite{Gorbar:2013qsa} it has been argued that due to this chiral shift a dynamical Hall effect is induced in a magnetic field in a Dirac metal which qualitatively is consitent with (\ref{eq:Hallcond}).

Finally there is one observation about the momentum and energy dissipations in the background of an AdS Schwarzschild black hole. As we already know, the DC conductivity dual to the Schwarzschild black hole is finite, which can be thought to be the result of zero density as can be seen from the $c=0$ limit in (\ref{conductivity}) that at zero density $\rho\to 0$, the DC conductivity for a translationally invariant system automatically becomes finite as the divergence term is proportional to the density and vanishes at zero density. Another way to understand this is that there is in fact a large amount of neutral degrees of freedom which dissipate the momentum of the charge carriers.  Analogous to this, we can see that in (\ref{conductivity}) the energy dissipation related term also vanishes at zero density, and with the same logic, this can be understood in another way, i.e. the energy of the charge carriers can be dissipated to the large amount of neutral degrees of freedom. This implies that the field theory dual to the AdS Schwarzschild black hole automatically has both energy and momentum dissipations for the charge carriers.

%%%%%%%%%%%%%%%%%%%%%%%%%%%%%%%%%%%%%%%%%
\section{A more realistic model: $U(1)_V\times U(1)_A$}
\label{sec4}
%%%%%%%%%%%%%%%%%%%%%%%%%%%%%%%%%%%%%%%%%

In Sec. \ref{sec2} and Sec. \ref{sec3}, we calculated the longitudinal electric conductivity for a chiral anomalous fluid with only one $U(1)$ current which in fact corresponds to the axial current. To be more realistic with the real electric current included, in this section we consider the case of two $U(1)$ currents, i.e. $U(1)_V \times U(1)_A$.

\subsection{Linear response}

Now we have two currents: $J^\mu$ which corresponds to $U(1)_V$ and $J_5^\mu$ which corresponds to $U(1)_A$. The conservation equations are now
\begin{eqnarray}
\partial_\mu T^{\mu\nu}&=&F^{\nu\alpha}J_{\alpha}\,,\\
\partial_\mu J^\mu&=& 0\,, \\
\partial_\mu J_5^\mu &=& c E^\mu B_\mu\,,
\end{eqnarray} when there are no dissipation terms.

The constituent equations are  \cite{Nielsen:1983rb,Sadofyev:2010pr}\footnote{\label{footnoteim} Note that $\mu=\frac{1}{2}(\mu_R+\mu_L),\mu_5=\frac{1}{2}(\mu_R-\mu_L); \rho=\rho_R+\rho_L, \rho_5=\rho_R-\rho_L.$ In this section, we use $\sigma$ to parametrize the quantum critical conductivity and $\Sigma$ for the total conductivity. }
\begin{eqnarray}
T^{\mu\nu}&=&\epsilon u^\mu u^\nu +p P^{\mu\nu}+\tau^{\mu\nu}\,,
\\
J^{\mu}&=& \rho u^\nu+ \nu^\mu\,,\\
J_5^{\mu}&=& \rho_5 u^\nu+ \nu_5^\mu
\end{eqnarray}
where in Landau frame
\begin{eqnarray}
\tau^{\mu\nu}&=&-\eta P^{\mu\alpha} P^{\nu\beta}(\partial_\alpha u_\beta+\partial_\beta u_\alpha)-(\zeta-\frac{2}{3}\eta)P^{\mu\nu}\partial_\alpha u^\alpha\,,\\
\nu^\mu&=&-\sigma TP^{\mu\nu}\partial_\nu\big(\frac{\mu}{T}\big)-\sigma_5 TP^{\mu\nu}\partial_\nu\big(\frac{\mu_5}{T}\big)+\sigma^{(E)} E^\mu+\sigma^{(V)} \omega^\mu+\sigma^{(B)} B^\mu\,, \\
\nu_5^\mu&=&-\sigma_5 TP^{\mu\nu}\partial_\nu\big(\frac{\mu}{T}\big)-\sigma TP^{\mu\nu}\partial_\nu\big(\frac{\mu_5}{T}\big)+\sigma_5^{(E)} E^\mu+\sigma_5^{(V)} \omega^\mu+\sigma_5^{(B)} B^\mu\,,
\end{eqnarray}
and 
\begin{equation}
E^\mu=F^{\mu\nu}u_\nu\,,~~~ B^\mu=\frac{1}{2}\epsilon^{\mu\nu\alpha\beta}u_\nu F_{\alpha\beta}\,,~~~\omega^\mu=\frac{1}{2}\epsilon^{\mu\nu\alpha\beta}u_\nu\partial_\alpha u_\beta\,.
\end{equation}
The coefficients are %\comment{cite1102.4334} 
\cite{Sadofyev:2010pr, {Kalaydzhyan:2011vx}}\footnote{For simplicity, we assume the gravitational anomaly constant $c_g=0$, thus we ignored the $T^2$ terms in these coefficients.}
%From $\partial_\mu (su^\mu)\geq 0$ \cite{Sadofyev:2010pr}, \comment{CHECK}
\begin{eqnarray} \sigma^{(E)}&=&\sigma(\mu, \mu_5, T)\,,~~
\sigma^{(B)}=c\mu_5\big(1-\frac{\mu\rho}{\epsilon+p}\big)\,, ~~
\sigma_5^{(B)}=c\mu\big(1-\frac{\mu_5\rho_5}{\epsilon+p}\big)\nonumber\\ \sigma_5^{(E)}&=&\sigma_5(\mu, \mu_5,T)\,,~~
\sigma^{(V)}=2c\mu\mu_5\big(1-\frac{\mu\rho}{\epsilon+p}\big)\,, ~~ \sigma_5^{(V)}=c\mu^2\big(1-\frac{2\mu_5\rho_5}{\epsilon+p}\big)\,.
%\sigma_V=c\big(\mu\mu_5-\frac{2}{3}\frac{\rho\mu\mu_5^2}{\epsilon+p}\big)
\end{eqnarray}
%up to $T^2$ terms. 
With two chemical potentials and two charges we now have the thermodynamic relations as %\comment{CHECK}
\begin{equation}
\epsilon+p=Ts+\mu\rho+\mu_5\rho_5\,,~~~dp=sdT+\rho d\mu+\rho_5 d\mu_5\,. 
\end{equation}
The nonzero quantities of the energy momentum tensor and currents are 
\begin{equation}
T^{00}=\epsilon\,,~~T^{ii}=p\,,~~ J^t=\rho\,,~~J^z=B\sigma^{(B)}\,,~~ J^t_5=\rho_5\,,~~J^z_5=B\sigma_5^{(B)}\,.
\end{equation}

We perform the same linear response calculations as in the one $U(1) $ current case. We assume that the system is in an equilibrium state characterized by $\mu$, $\mu_5$ and the temperature $T$.  We perturb the system by
\begin{eqnarray}
\mu(\vec{x},t)&=&\mu+\delta \mu(\vec{x},t)\,,\\
\mu_5(\vec{x},t)&=&\mu_5+\delta \mu_5(\vec{x},t)\,,\\
T(\vec{x},t)&=&T+\delta T(\vec{x},t)\,,\\
u^\mu(\vec{x},t)&=&(1,\delta u_i(\vec{x},t))\,,
\end{eqnarray}
and $\delta F^{0i}=-\delta F^{i0}$ in the $U(1)_V$ sector. We also have a background magnetic field in the $U(1)_V$ sector $F_{12}=-F_{21}=B,$ and $ E^\mu=0.$

To linear order we have
\begin{eqnarray}
\delta T^{00}&=&\delta \epsilon\,,\nonumber\\
\delta T^{0i}&=&(\epsilon+p)\delta u^i\,,\nonumber\\
\delta T^{ij}&=&\delta p g^{ij}-\eta\big(\partial^i \delta u^j+\partial^j \delta u^i-\frac{2}{3}g^{ij}\partial_k\delta u_k\big)-\zeta g^{ij}\partial_k\delta u_k\,,\nonumber\\
\delta J^t&=&\delta \rho+\sigma^{(B)} B\delta u_z\,,\nonumber\\
\delta J^x&=&\rho\delta u_x+\sigma^{(E)}\big(\delta F^{0x}+B\delta u_y\big)-\sigma T\partial_x\bigg(\delta \frac{\mu}{T}\bigg)-\sigma_5 T\partial_x\bigg(\delta \frac{\mu_5}{T}\bigg)+\frac{1}{2}\sigma^{(V)}\big(\partial_y \delta u_z-\partial_z\delta u_y\big)\,,\nonumber\\
\delta J^y&=&\rho\delta u_y+\sigma^{(E)}\big(\delta F^{0y}-B\delta u_x\big)-\sigma T\partial_y\bigg(\delta \frac{\mu}{T}\bigg)-\sigma_5 T\partial_y\bigg(\delta \frac{\mu_5}{T}\bigg)-\frac{1}{2}\sigma^{(V)}\big(\partial_x \delta u_z-\partial_z\delta u_x\big)\,,\nonumber\\
\delta J^z&=&\rho\delta u_z+\sigma^{(E)}\delta E_z-\sigma T\partial_z\bigg(\delta \frac{\mu}{T}\bigg)-\sigma_5 T\partial_z\bigg(\delta \frac{\mu_5}{T}\bigg)+\frac{1}{2}\sigma^{(V)}\big(\partial_x \delta u_y-\partial_y\delta u_x\big)+\delta\sigma^{(B)}B\,,\nonumber\\
\delta J_5^t&=&\delta \rho_5+\sigma_5^{(B)} B\delta u_z\,,\nonumber\\
\delta J_5^x&=&\rho_5\delta u_x+\sigma_5^{(E)}\big(\delta F^{0x}+B\delta u_y\big)-\sigma T\partial_x\bigg(\delta \frac{\mu_5}{T}\bigg)-\sigma_5 T\partial_x\bigg(\delta \frac{\mu}{T}\bigg)+\frac{1}{2}\sigma_5^{(V)}\big(\partial_y \delta u_z-\partial_z\delta u_y\big)\,,\nonumber\\
\delta J_5^y&=&\rho_5\delta u_y+\sigma_5^{(E)}\big(\delta F^{0y} -B\delta u_x\big)-\sigma T\partial_y\bigg(\delta \frac{\mu_5}{T}\bigg)-\sigma_5 T\partial_x\bigg(\delta \frac{\mu}{T}\bigg)-\frac{1}{2}\sigma_5^{(V)}\big(\partial_x \delta u_z-\partial_z\delta u_x\big)\,,\nonumber\\
\delta J_5^z&=&\rho_5\delta u_z+\sigma_5^{(E)}\delta E_z-\sigma T\partial_z\bigg(\delta \frac{\mu_5}{T}\bigg)-\sigma_5 T\partial_x\bigg(\delta \frac{\mu}{T}\bigg)+\frac{1}{2}\sigma_5^{(V)}\big(\partial_x \delta u_y-\partial_y\delta u_x\big)+\delta\sigma_5^{(B)}B\,.\nonumber
\end{eqnarray}
We consider the following conservation equations at linear order in 
$\delta J^{\mu},$ $\delta J^{\mu}_5$, $\delta T_{\mu\nu} $ \cite{nernsteffect} again with all the possible dissipation terms except for the $U(1)_V$ current which remains conserved
\begin{eqnarray}
\partial_{\mu}\delta T^{\mu 0}&=&\delta F^{0\mu} J_\mu+\frac{1}{\tau_e}\delta T^{\mu 0}u_{\mu}\,,\\
\partial_{\mu}\delta T^{\mu i}&=&\rho\delta E^i+F^{i\lambda}\delta J_{\lambda}+\frac{1}{\tau_m}\delta T^{\mu i}u_{\mu}\,,\\
\partial_{\mu}\delta J^\mu&=&0\,,\\
\partial_{\mu}\delta J_5^{\mu}&=&c\delta E^\mu B_\mu+\frac{1}{\tau_c}\delta J_5^\mu u_\mu\,.
\end{eqnarray}

After substituting the currents and stress energy tensor at the linear level into the equations above, we get the following conservation equations
\begin{eqnarray}
\big(\partial_t+\frac{1}{\tau_e}\big)\delta\epsilon+\partial_i\big[(\epsilon+p)\delta u_i\big]-\sigma^{(B)} B \delta E_z&=&0\,;\nonumber\\
\big(\partial_t+\frac{1}{\tau_m}\big)\big[(\epsilon+p)\delta u_x\big]+\partial_x\delta p-\eta \big(\partial_j^2\delta u_x+\frac{1}{3}\partial_x\partial_j\delta u_j\big)
-\zeta\partial_x\partial_j\delta u_j
=\rho\delta F^{0x}&&\nonumber\\
+
B\bigg[\rho\delta u_y+\sigma^{(E)}\big(\delta F^{0y}-B\delta u_x\big)-\sigma T\partial_y\bigg(\delta \frac{\mu}{T}\bigg)-\sigma_5 T\partial_y\bigg(\delta \frac{\mu_5}{T}\bigg)-\frac{1}{2}\sigma^{(V)}\big(\partial_x \delta u_z-\partial_z\delta u_x\big)\bigg];&&\nonumber\\
\big(\partial_t+\frac{1}{\tau_m}\big)\big[(\epsilon+p)\delta u_y\big]+\partial_y\delta p-\eta \big(\partial_j^2\delta u_y+\frac{1}{3}\partial_y\partial_j\delta u_j\big)
-\zeta\partial_y\partial_j\delta u_j
=\rho\delta F^{0y}&&\nonumber\\
-B\bigg[\rho\delta u_x+\sigma^{(E)}\big(\delta F^{0x}+B\delta u_y\big)-\sigma T\partial_x\bigg(\delta \frac{\mu}{T}\bigg)-\sigma_5 T\partial_x\bigg(\delta \frac{\mu_5}{T}\bigg)+\frac{1}{2}\sigma^{(V)}\big(\partial_y \delta u_z-\partial_z\delta u_y\big)\bigg]\,;&&\nonumber\\
\big(\partial_t+\frac{1}{\tau_m}\big)\big[(\epsilon+p)\delta u_z\big]+\partial_z\delta p-\eta \big(\partial_j^2\delta u_z+\frac{1}{3}\partial_z\partial_j\delta u_j\big)
-\zeta\partial_z\partial_j\delta u_j-\rho\delta E_z
&=&0\,;\nonumber\\
%\big(\partial_t+\frac{1}{\tau_2}\big)
\partial_t\big[\delta \rho+\sigma_B B\delta u_z\big]+\partial_i(\rho \delta u_i+\sigma^{(E)}\delta E_i)-\sigma T\partial_i^2\big(\delta \frac{\mu}{T}\big)-\sigma_5 T\partial_i^2\big(\delta \frac{\mu_5}{T}\big) 
&&\nonumber\\
+\sigma^{(E)} B(\partial_x\delta u_y-\partial_y\delta u_x)+\partial_z\delta\sigma^{(B)} B &=&0\,;\nonumber\\
\big(\partial_t+\frac{1}{\tau_c}\big)\big[\delta \rho_5+\sigma_5^{(B)} B\delta u_z\big]+\partial_i(\rho_5 \delta u_i+\sigma_5^{(E)}\delta E_i)-\sigma T\partial_i^2\big(\delta \frac{\mu_5}{T}\big)
-\sigma_5 T\partial_i^2\big(\delta \frac{\mu}{T}\big)
&&\nonumber\\
+\sigma_5^{(E)} B(\partial_x\delta u_y-\partial_y\delta u_x)
+\partial_z\delta\sigma_5^{(B)} B-cB\delta E_z &=&0\,.\nonumber
\end{eqnarray}

Laplace transform in the time direction and we get
\begin{eqnarray}
\omega_e\delta\epsilon-i\delta \epsilon^{(0)}+i(\epsilon+p)\partial_i\delta u_i
-i\sigma^{(B)} B \delta E_z&=&0\,,\nonumber\\
(\epsilon+p)(\omega_m\delta u_x-i\delta u_x^{(0)})+i\partial_x\delta p-i\eta(\partial_j^2\delta u_x+\frac{1}{3}\partial_x\partial_j\delta u_j)-i\zeta\partial_x\partial_j\delta u_j
-iB\delta J_y-i\rho\delta F^{0x}&=&0\,,\nonumber\\
(\epsilon+p)(\omega_m\delta u_y-i\delta u_y^{(0)})+i\partial_y\delta p-i\eta(\partial_j^2\delta u_y+\frac{1}{3}\partial_y\partial_j\delta u_j)-i\zeta\partial_y\partial_j\delta u_j+iB\delta J_x-i\rho\delta F^{0y}&=&0\,,\nonumber\\
(\epsilon+p)(\omega_m\delta u_z-i\delta u_z^{(0)})+i\partial_z\delta p-i\eta(\partial_j^2\delta u_z+\frac{1}{3}\partial_z\partial_j\delta u_j)-i\zeta\partial_z\partial_j\delta u_j-i\rho\delta E_z&=&0\,,\nonumber\\
%\big(\omega\delta\rho-i\delta \rho^{(0)}\big)+\sigma_B B\big(\omega_2\delta u_z-i\delta u_3^{(0)}\big)
\omega\big(\delta\rho+\sigma^{(B)} B\delta u_z\big)
+i\partial_i(\rho \delta u_i+\sigma^{(E)}\delta E_i)-i\sigma T\partial_i^2\big(\delta \frac{\mu}{T}\big)-i\sigma_5 T\partial_i^2\big(\delta \frac{\mu_5}{T}\big)
&&\nonumber\\
+i\sigma^{(E)} B(\partial_x\delta uy-\partial_y\delta u_x)
+i\partial_z\delta\sigma_B B%-ic B\delta E_3
&=&0\,,\nonumber\\
\big(\omega_c\delta\rho_5-i\delta \rho_5^{(0)}\big)+
\sigma_5^{(B)} B\big(\omega_c\delta u_z-i\delta u_z^{(0)}\big)+i\partial_i(\rho_5 \delta u_i+\sigma_5^{(E)}\delta E_i)-i\sigma T\partial_i^2\big(\delta \frac{\mu_5}{T}\big)
&&\nonumber\\
-i\sigma_5 T\partial_i^2\big(\delta \frac{\mu}{T}\big)
+i\sigma_5^{(E)} B(\partial_x\delta u_y-\partial_y\delta u_x)
+i\partial_z\delta\sigma_5^{(B)} B-ic B\delta E_z
&=&0\,,\nonumber
\end{eqnarray}
where $$\omega_e\equiv \omega+\frac{i}{\tau_e}\,, ~~\omega_m\equiv \omega+\frac{i}{\tau_m}\,, ~~\omega_c\equiv \omega+\frac{i}{\tau_c}\,.$$

As $\delta F^{0\mu}$ is an external field, we can choose it to be $\delta F^{0\mu}(t,x^i)=\delta F^{0\mu}e^{-i \omega t+i k_i x^i}.$ After a Laplace transformation in the time direction and a Fourier transformation in the spatial direction and taking the limit $k\to 0$ we have $\delta E_z=\delta E_z^{(0)}$. The equations become
\begin{eqnarray}
\omega_e\delta\epsilon-i\delta \epsilon^{(0)}%+(\epsilon+p)\partial_z \delta u_z
-i\sigma^{(B)} B \delta E_z&=&0\,,\\
(\epsilon+p)(\omega_m\delta u_x-i\delta u_x^{(0)})%+i \partial_x \delta p
-i\rho\delta F^{0x} %\nonumber\\  
-iB\big(\rho\delta u_y+\sigma^{(E)}(\delta F^{0y}%-\partial_y \delta \mu
- B\delta u_x
%+\mu\frac{\partial_y \delta T}{T}
)%-\frac{1}{2}\sigma^{(V)}\partial_x\delta u_z
\big)
&=&0\,,\\
(\epsilon+p)(\omega_m\delta u_y-i\delta u_y^{(0)})%+i\partial_y \delta p
-i\rho\delta F^{0y}%\nonumber\\
+iB\big(\rho\delta u_x+\sigma^{(E)}(\delta F^{0x}%-\partial_x \delta \mu
+ B\delta u_y%+\mu\frac{\partial_x \delta T}{T}
%+\frac{1}{2}\sigma^{(V)}\partial_y \delta u_z
)\big)&=&0\,,\\
(\epsilon+p)(\omega_m\delta u_z-i\delta u_z^{(0)})-i\rho\delta E_z%+i\partial_z \delta p
&=&0\,,\\
\omega\big(\delta\rho+
\sigma^{(B)} B \delta u_z\big)&=&0\,,\\
\omega_c\delta\rho_5-i\delta \rho_5^{(0)}+
\sigma_5^{(B)} B\big(\omega_c\delta u_z-i\delta u_z^{(0)}\big)-ic B\delta E_z&=&0\,.
\end{eqnarray}

Solving $\delta\mu,\delta \rho,\delta u_z$ in terms of 
$\delta\mu^{(0)},\delta \rho^{(0)},\delta u_z^{(0)},\delta E_z^{(0)}$ 
and using 
\begin{eqnarray}
\delta\epsilon&\equiv&e_5\delta \mu_5+e_1\delta\mu+e_2\delta T=\big(\frac{\partial\epsilon}{\partial \mu_5}\big)\Big{|}_{T,\mu}\delta \mu_5+\big(\frac{\partial\epsilon}{\partial \mu}\big)\Big{|}_{T,\mu_5}\delta \mu+\big(\frac{\partial\epsilon}{\partial T}\big)\Big{|}_{\mu,\mu_5}\delta T\,,\\
\delta\rho&\equiv&f_5\delta \mu_5+f_1\delta\mu+f_2\delta T=\big(\frac{\partial\rho}{\partial \mu_5}\big)\Big{|}_{T,\mu}\delta \mu_5+\big(\frac{\partial\rho}{\partial \mu}\big)\Big{|}_{T,\mu_5}\delta \mu+\big(\frac{\partial\rho}{\partial T}\big)\Big{|}_{\mu,\mu_5}\delta T\,,\\
\delta\rho_5&\equiv&s_5\delta \mu_5+s_1\delta\mu+s_2\delta T=\big(\frac{\partial\rho_5}{\partial \mu_5}\big)\Big{|}_{T,\mu}\delta \mu_5+\big(\frac{\partial\rho_5}{\partial \mu}\big)\Big{|}_{T,\mu_5}\delta \mu+\big(\frac{\partial\rho_5}{\partial T}\big)\Big{|}_{\mu,\mu_5}\delta T\,,\\
\delta p&=&\rho_5\delta \mu_5+\rho\delta \mu+s \delta T\,,
\end{eqnarray}
we have 
\begin{eqnarray}
\delta u_z&=&\frac{\rho}{\epsilon+p}\frac{i}{\omega_m}\delta E_z^{(0)}+\dots\nonumber\\
\delta \mu&=&\frac{B}{D}\bigg[\sigma^{(B)}(f_5 s_2-f_2 s_5)\frac{i}{\omega_e}
+
\frac{\rho[\sigma^{(B)}(e_5 s_2-e_2 s_5)-\sigma_5^{(B)}(e_5 f_2-e_2 f_5)]}{\epsilon+p}\frac{i}{\omega_m}\nonumber\\&&~~~+c(e_5 f_2-e_2 f_5)\frac{i}{\omega_c}
\bigg]\delta E_z^{(0)}+\dots\nonumber\\
\delta \mu_5&=&\frac{B}{D}\bigg[\sigma^{(B)}(f_2 s_1-f_1 s_2)\frac{i}{\omega_e}
+
\frac{\rho[\sigma^{(B)}(e_2 s_1-e_1 s_2)-\sigma_5^{(B)}(e_2 f_1-e_2 f_1)]}{\epsilon+p}\frac{i}{\omega_m}
\nonumber\\&&~~~+c(e_2 f_1-e_1 f_2)\frac{i}{\omega_c}
\bigg]\delta E_z^{(0)}+\dots\nonumber\\
\delta T&=&-\frac{B}{D}\bigg[\sigma^{(B)}(f_5 s_1-f_1 s_5)\frac{i}{\omega_e}
+
\frac{\rho[\sigma^{(B)}(e_5 s_1-e_1 s_5)-\sigma_5^{(B)}(e_5 f_1-e_1 f_5)]}{\epsilon+p}\frac{i}{\omega_m}\nonumber\\&&~~~+c(e_5 f_1-e_1 f_5)\frac{i}{\omega_c}
\bigg]\delta E_z^{(0)}+\dots\nonumber
\end{eqnarray}
with 
\begin{equation}
D\equiv \text{det} \begin{pmatrix} e_5 & e_2 & e_1 \\   f_5 & f_2 & f_1 \\
s_5 & s_2 & s_1 \end{pmatrix},
\end{equation}where ``\dots" denote terms unrelated to $\delta E_z^{(0)}$. Here we only focus on the longitudinal conductivity with vanishing initial values for all the other perturbations except $\delta E_z^{(0)}$.
%Equation (3.34) in 0706.3215 is not a little different because of the change in the conservation of current.

From \begin{equation}\delta J^z=\rho\delta u_z+\sigma^{(E)}\delta E_z-\sigma T\partial_z\bigg(\delta \frac{\mu}{T}\bigg)+\frac{1}{2}\sigma^{(V)}\big(\partial_x \delta u_y-\partial_y\delta u_x\big)+\delta\sigma^{(B)}B\,,\end{equation}
we get (in the $k\to 0$ limit) %\comment{CHECK}
\begin{eqnarray}
\delta J^z&=&\rho\delta u_z+\sigma^{(E)}\delta E_z+Bc\bigg(1-\frac{\mu\rho+\mu\mu_5f_5}{(\epsilon+p)}+\frac{\rho\mu\mu_5(e_5+\rho_5)}{(\epsilon+p)^2}\bigg)\delta \mu_5
\nonumber\\&&~~~+Bc\bigg(-\frac{\mu_5\rho+\mu\mu_5f_1}{(\epsilon+p)}+\frac{\rho\mu\mu_5(e_1+\rho)}{(\epsilon+p)^2}\bigg)\delta \mu \nonumber\\
&&~~~+Bc\bigg(-\frac{\mu\mu_5f_2}{(\epsilon+p)}+\frac{\rho\mu\mu_5(e_2+s)}{(\epsilon+p)^2}\bigg)\delta T\\
&=&{\Sigma} \delta E_z^{(0)}+\dots
\end{eqnarray}
 with 
\begin{equation}\label{conductivity2}
{{\Sigma}}=\sigma^{(E)}+\frac{i}{\omega+\frac{i}{\tau_e}}\frac{B^2c\sigma^{(B)}}{D}K_0+\frac{i}{\omega+\frac{i}{\tau_m}}\frac{\rho}{\epsilon+p}\bigg[\rho-
\frac{B^2 c}{D} K_1\bigg]+\frac{i}{\omega+\frac{i}{\tau_c}}\frac{B^2c^2}{D}K_2\,,
\end{equation}
where
\begin{eqnarray}
K_0&=&(f_2s_1-f_1s_2)%+(f_5s_2-f_2s_5)
-\frac{\rho}{(\epsilon+p)}\big[(f_2s_1-f_1s_2)\mu+(f_5s_2-f_2s_5)\mu_5\big]\nonumber\\&&+\frac{\mu\mu_5\rho}{(\epsilon+p)^2}\big[D-s(f_5s_1-f_1s_5)+\rho(f_5s_2-f_2s_5)+\rho_5(f_2s_1-f_1s_2)\big]\\
K_1&=&\sigma^{(B)}\bigg[(e_1s_2-e_2s_1)%-(e_5s_2-e_2s_5)
-\frac{1}{(\epsilon+p)}\big[\mu\mu_5D-\mu_5\rho(e_5s_2-e_2s_5)\nonumber\\&&~~~-\mu\rho(e_2s_1-e_1s_2)\big]-\frac{\mu\mu_5\rho}{(\epsilon+p)^2}\big[-s(e_5s_1-e_1s_5)+\rho (e_5 s_2-e_2 s_5)\nonumber\\
&&+\rho_5(e_2 s_1-e_1 s_2)\big]\bigg]+\sigma_5^{(B)} K_2\\
%\bigg[(e_2f_1-e_1f_2)+(e_5f_2-e_2f_5)-\frac{\rho}{2(\epsilon+p)}\big[\mu_5(e_5f_2-e_2f_5)\nonumber\\&&~~~+\mu(e_2f_1-e_1f_2)\big]-\frac{\mu\mu_5\rho}{2(\epsilon+p)^2}\big[s(e_5f_1-e_1f_5))
%-\rho (e_5 f_2-e_2 f_5)-\rho_5(e_2 f_1-e_1 f_2)\big]\bigg],\\ 
K_2&=&(e_2f_1-e_1f_2)%+(e_5f_2-e_2f_5)
-\frac{\rho}{(\epsilon+p)}\left[%2\mu\mu_5f_2(e_1f_5-e_5f_1)
\mu(e_2f_1-e_1f_2)+\mu_5(e_5f_2-e_2f_5)\right]
\nonumber\\&&
+\frac{\mu\mu_5\rho}{(\epsilon+p)^2}\left[-s(e_5f_1-e_1f_5)+\rho (e_5 f_2-e_2 f_5)+\rho_5(e_2 f_1-e_1 f_2)\right].\end{eqnarray}

The result is very complicated with lots of thermodynamic quantities that are not universal. In certain limits, the result can be simplified to very simple forms.
\begin{itemize}
%$\bullet$ 
\item case I: $\rho_5=\mu_5=0.$ In this case we have $s_1=s_2=0,$ and

\begin{equation}
\Sigma=\sigma^{(E)}+\frac{i}{\omega+\frac{i}{\tau_m}}\frac{\rho}{\epsilon+p}\bigg[\rho-\frac{B^2c^2\mu Ts}{s_5(\epsilon+p)}\bigg]+\frac{i}{\omega+\frac{i}{\tau_c}}\frac{Ts}{\epsilon+p}\frac{B^2 c^2}{ s_5}\,.
\end{equation}In this case, energy dissipations are not necessary for a finite result while momentum and charge dissipations are still required. This is a very interesting limit. It is clear from (\ref{conductivity2}) that energy dissipations are only needed when there is a non vanishing axial chemical potential. In this limit, the anomaly related dissipations are the charge and momentum dissipations, which  means that in this limit, the inter-valley scattering would have the effect of only dissipating charge and momentum, while not energy.

\item  case II: $\rho=\mu=0.$ We have $f_5=f_2=0,$ and
\begin{equation}
\Sigma=\sigma^{(E)}-\frac{i}{\omega+\frac{i}{\tau_e}}\frac{B^2c^2\mu_5 s_2}{(\epsilon+p)(e_2s_5-e_5s_2)}-\frac{i}{\omega+\frac{i}{\tau_c}}\frac{B^2 c^2e_2}{ e_5s_2-e_2s_5}\,. 
\end{equation}In this case, momentum dissipation is not necessary while energy and charge dissipations are needed for a finite result. This is because momentum dissipation is always associated with finite charge density.

\item  case III: $\rho=\rho_5=0, \mu=\mu_5=0.$ 
Using the fact $f_5=f_2=s_1=s_2=0,$ we have
\begin{equation}\label{con-twou1}
\Sigma=\sigma^{(E)}+\frac{i}{\omega+\frac{i}{\tau_c}}\frac{B^2c^2}{s_5}\,.
\end{equation}
In this double zero density limit, only charge dissipations are needed for a finite result.

\end{itemize}

Another interesting quantity is the axial longitudinal conductivity $\Sigma_{5}$ which is defined as the response of $J_{Az}$ to the electric field $E_z$. In the $k\to 0$ limit we have
\begin{equation}\delta J_5^z=\rho_5\delta u_z+\sigma_5^{(E)}\delta E_z+\delta\sigma_5^{(B)}B\,,\end{equation}
then we get %\comment{CHECK}
\begin{eqnarray}
\delta J_5^z&=&\rho_5\delta u_z+\sigma_5^{(E)}\delta E_z+Bc\bigg(-\frac{\mu\rho_5+\mu\mu_5s_5}{(\epsilon+p)}+\frac{\rho_5\mu\mu_5(e_5+\rho_5)}{(\epsilon+p)^2}\bigg)\delta \mu_5
\nonumber\\&&~~~+Bc\bigg(1-\frac{\mu_5\rho_5+\mu\mu_5s_1}{(\epsilon+p)}+\frac{\rho_5\mu\mu_5(e_1+\rho)}{(\epsilon+p)^2}\bigg)\delta \mu \nonumber\\
&&~~~+Bc\bigg(-\frac{\mu\mu_5s_2}{(\epsilon+p)}+\frac{\rho_5\mu\mu_5(e_2+s)}{(\epsilon+p)^2}\bigg)\delta T\\
&=&{\Sigma}_5 \delta E_z^{(0)}+\dots
\end{eqnarray}
 with 
\begin{equation}\label{conductivity5}
\Sigma_5=\sigma_5^{(E)}+\frac{i}{\omega+\frac{i}{\tau_e}}\frac{B^2c\sigma^{(B)}}{D}W_0+\frac{i}{\omega+\frac{i}{\tau_m}}\frac{\rho}{\epsilon+p}\bigg[\rho_5-
\frac{B^2 c}{D} W_1\bigg]+\frac{i}{\omega+\frac{i}{\tau_c}}\frac{B^2c^2}{D}W_2\,,
\end{equation}
where
\begin{eqnarray}
W_0&=&(f_5s_2-f_2s_5)
-\frac{\rho_5}{(\epsilon+p)}\big[(f_2s_1-f_1s_2)\mu+(f_5s_2-f_2s_5)\mu_5\big]\nonumber\\&&+\frac{\mu\mu_5\rho_5}{(\epsilon+p)^2}\big[D-s(f_5s_1-f_1s_5)+\rho(f_5s_2-f_2s_5)+\rho_5(f_2s_1-f_1s_2)\big]\,,\\
W_1&=&\sigma^{(B)}\bigg[(e_2s_5-e_5s_2)%-(e_5s_2-e_2s_5)
+\frac{\rho_5}{(\epsilon+p)}\big[\mu(e_2s_1-e_1s_2)+\mu_5(e_5s_2-e_2s_5)\big]\nonumber\\&&~~~-\frac{\mu\mu_5\rho_5}{(\epsilon+p)^2}\big[-s(e_5s_1-e_1s_5)+\rho (e_5 s_2-e_2 s_5)
+\rho_5(e_2 s_1-e_1 s_2)\big]\bigg]
\nonumber\\&&+\sigma_5^{(B)} W_2\,,\\
%\bigg[(e_2f_1-e_1f_2)+(e_5f_2-e_2f_5)-\frac{\rho}{2(\epsilon+p)}\big[\mu_5(e_5f_2-e_2f_5)\nonumber\\&&~~~+\mu(e_2f_1-e_1f_2)\big]-\frac{\mu\mu_5\rho}{2(\epsilon+p)^2}\big[s(e_5f_1-e_1f_5))
%-\rho (e_5 f_2-e_2 f_5)-\rho_5(e_2 f_1-e_1 f_2)\big]\bigg],\\ 
W_2&=&(e_5f_2-e_2f_5)
-\frac{1}{(\epsilon+p)}\left[%2\mu\mu_5f_2(e_1f_5-e_5f_1)
\mu\mu_5 D+\mu\rho_5(e_2f_1-e_1f_2)+\mu_5\rho_5(e_5f_2-e_2f_5)
\right]
\nonumber\\&&
+\frac{\mu\mu_5\rho_5}{(\epsilon+p)^2}\left[-s(e_5f_1-e_1f_5)+\rho (e_5 f_2-e_2 f_5)+\rho_5(e_2 f_1-e_1 f_2)\right].\end{eqnarray}

We can also simplify the results above in certain limits:
\begin{itemize}
%$\bullet$ 
\item In the limit $B=0$, it reduces to a simple result \begin{equation}\Sigma_5=\sigma_5^{(E)}+\frac{i}{\omega+\frac{i}{\tau_m}}\frac{\rho\rho_5}{\epsilon+p}.\end{equation}
\item In the limit $\rho_5=\mu_5=0.$ In this case we have %$s_1=s_2=0,$
\begin{equation}
\Sigma_5=\sigma_5^{(E)}+\frac{i}{\omega+\frac{i}{\tau_m}}\frac{\rho}{\epsilon+p}\bigg[-\frac{B^2c^2\mu (e_5f_2-e_2f_5)}{s_5(e_2f_1-e_1f_2)}\bigg]+\frac{i}{\omega+\frac{i}{\tau_c}}\frac{B^2 c^2(e_5f_2-e_2f_5)}{ s_5(e_2f_1-e_1f_2)}\,,
\end{equation}where only momentum and charge dissipations are needed.

\item In the limit $\rho=\mu=0.$ We have %$e_1=f_5=f_2=s_1=0.$
$
\Sigma_5=\sigma_5^{(E)} 
$ and is automatically finite without any dissipation terms.

\item  When $\rho=\rho_5=0, \mu=\mu_5=0,$ 
we have
$
\Sigma_5=\sigma_5^{(E)}.
$

\end{itemize}

%%%%%%%%%%%%%%%%%%%%%%%%%
\subsection{Holographic calculations}
%%%%%%%%%%%%%%%%%%%%%%%%%

Now we apply this formula to the holographic system in the probe limit. The holographic $U(1)_V\times U(1)_A$ model was proposed in \cite{Gynther:2010ed} with the action
\begin{equation}\
S=\int d^5x \sqrt{-g}\left[-\frac{1}{4}F_A^2-\frac{1}{4}F_V^2+\frac{\alpha}{3}\epsilon^{\mu\nu\rho\sigma\tau}A_\mu \big(F^A_{\nu\rho} F^A_{\sigma\tau}+3F^V_{\nu\rho} F^V_{\sigma\tau}\big)\right],
\end{equation}where subscript $V$ denotes the vector sector and $A$ denotes the axial sector. The equations of motion for the two gauge fields are
\begin{eqnarray}
\nabla_\nu F^{\nu\mu}_A+\alpha\epsilon^{\mu\alpha\beta\rho\sigma}\left(F^A_{\alpha\beta}F^A_{\rho\sigma}+F^V_{\alpha\beta}F^V_{\rho\sigma}
\right)&=&0\,,\\
\nabla_\nu F^{\nu\mu}_V+2\alpha\epsilon^{\mu\alpha\beta\rho\sigma} F^A_{\alpha\beta}F^V_{\rho\sigma}&=&0\,.
\end{eqnarray} 
To have a magnetic field in the background of the $U(1)_V$ sector, we assume $A^A_\mu=\big(a(r),0,0,c(r),0\big),$ $A^V_\mu=\big(a_2(r),By,0,c_2(r),0\big)$ and it is easy to check that $a, c_2$ and $a_2, c$ satisfy exactly the same equations as (\ref{bgeqnataz}). Thus
%Redefine $A^\mu_L=\frac{1}{2}(V^\mu-A^\mu), A^\mu_R=\frac{1}{2}(V^\mu+A^\mu),$ we obtaine two copies of independent Chern-Simons term with opposite $c$. From (\ref{rhomuinsch}) and relations in footnote \ref{footnoteim}, 
we have 
\begin{equation}
f_1=s_5=4 r_0^2 \frac{\Gamma\big[\frac{5-\sqrt{1-(8B\tilde{\alpha})^2}}{4}\big]\Gamma\big[\frac{5+\sqrt{1-(8B\tilde{\alpha})^2}}{4}\big]}{\Gamma\big[\frac{3-\sqrt{1-(8B\tilde{\alpha})^2}}{4}\big]\Gamma\big[\frac{3+\sqrt{1-(8B\tilde{\alpha})^2}}{4}\big]}\,,~~~~ f_5=s_1=0\,.
\end{equation}
Substituting these into the formulas and we get
\begin{equation}
\Sigma=\sigma_E+\frac{i}{\omega_c}\frac{B^2c^2}{4\pi^2 T^2}\frac{\Gamma\big[\frac{3-\sqrt{1-(8B\tilde{\alpha})^2}}{4}\big]\Gamma\big[\frac{3+\sqrt{1-(8B\tilde{\alpha})^2}}{4}\big]}{\Gamma\big[\frac{5-\sqrt{1-(8B\tilde{\alpha})^2}}{4}\big]\Gamma\big[\frac{5+\sqrt{1-(8B\tilde{\alpha})^2}}{4}\big]}\,.
\end{equation}
This is the prediction for the probe holographic system from the hydrodynamic linear response theory. We can compare this with the results directly from holography.

The conductivity of $U(1)_V\times U(1)_A$ can be computed from Kubo formula via holographic approach. It turns out the fluctuations of $(\delta A_t^V, \delta A_z^A)$ and $(\delta A_t^A, \delta A_z^V)$ form the same equations as a single U(1) case, thus it is easy to check that the holographic result for $\Sigma$ matches exactly our result from hydrodynamics. This means that in the $U(1)_V\times U(1)_A$ case we get the same behavior of the DC longitudinal magnetoconductivity as in previous section though with different physical meaning. The result in this $U(1)_V\times U(1)_A$ case is the real magnetoconductivity which should be compared with the experiments. As the result for $\Sigma$ is exactly the same as that found in Sec. \ref{sec3}, all the discussions in Sec. \ref{sec3} are still valid, i.e. holography can naturally realize the negative magnetoresistivity in an intermediate regime of $B$ and provides a new explanation of the decrease of the magnetoconductivity as a function of $B$ at small $B$ as found in experiments. For $\Sigma_5$ as $\delta A_z^V$ and $\delta A_z^A$ are not coupled, we have $\Sigma_5=0$ in the probe limit of the holographic system.

%%%%%%%%%%%%%%%%%%%%%%%%%%%%%%%%%%%%%%%%%
\section{Conclusion and discussion}
\label{sec-con}
%%%%%%%%%%%%%%%%%%%%%%%%%%%%%%%%%%%%%%%%%

In this paper we have considered the behavior of the longitudinal electric conductivity with a background magnetic field in a chiral fluid. When there is an electric field parallel to the magnetic field, an  anomaly related infinite DC longitudinal conductivity will arise due to infinite increase of the chemical potential and temperature under the external electric field, which is caused by the chiral anomaly. We calculated the longitudinal conductivity in the hydrodynamic limit at the linear response level for both the cases with one $U(1)$ current and  $U(1)_V \times U(1)_A$ currents.  The results show that the infinite conductivity can only become finite by including all the three kinds of possible dissipation terms: momentum dissipation, energy dissipation and charge dissipation. Even at zero density, there is still one infinite term left which can only be dissipated by the charge dissipation term.

We applied the formula of the magnetoconductivity which we got from hydrodynamic calculations to a simple holographic system in the probe limit and confirmed that it matches with the result from the Kubo formula in the holographic side.  The holographic result has a nontrivial dependence on the background magnetic field. In an intermediate regime of $B$, it grows linearly in $B$ which corresponds to the behavior of negative magnetoresistivity  and agrees both with the results obtained previously in \cite{Nielsen:1983rb, {sonspivak}} using the weakly coupled kinetic theory and those found in experiments. Our holographic result provides also a possible new explanation for
the decrease in the magnetoconductivity observed in experiment at small $B$. Indeed the quantum critical conductivity $\sigma_{Ez}$ along
the direction of the magnetic field is strongly quenched by $B$. Combined with the chiral magnetic effect and a small enough charge relaxation time
this leads to a dip in the magnetoconductivity, qualitatively similar to what has been observed in \cite{weyl-exp}.
In contrast the transverse DC conductivity is not affected by the magnetic field in the probe limit.
This conclusion will change once the system is at finite density with backreaction, and our next step is to study the chiral anomalous system holographically at finite density and see what would be the behavior of both the longitudinal and transverse magnetoconductivities, especially if there will be a cusp-like behavior at small $B$.

In our holographic calculations, we have not included dissipation effects. It would be very interesting to test the dissipation effects holographically.
Recently there has been a lot of work in including momentum dissipation in holography. These include the lattice construction which breaks the translational symmetry explicitly (e.g. \cite{Horowitz:2012ky,{Liu:2012tr},{Donos:2012js},{Donos:2013eha},{Andrade:2013gsa}}) and massive gravity which breaks the diffeomorphism symmetry in the bulk (e.g. \cite{Vegh:2013sk,{Davison:2013jba},{Davison:2013txa}}). Besides momentum dissipations, we also need to include energy and charge dissipations.

In \cite{Jimenez-Alba:2014iia}, a bulk massive gauge theory was studied in the chiral anomalous fluid  (see also \cite{Gursoy:2014ela}). The massive gauge theory breaks the $U(1)$ gauge symmetry in the bulk and leads to charge dissipation for the boundary theory. In a follow up paper, we plan to study the fluid/gravity analysis of this theory  (similar to \cite{Erdmenger:2008rm,{Banerjee:2008th}}) to get the charge relaxation time from the hydrodynamic modes \cite{inprogress}.

The holographic energy dissipation effects have not been considered so far. As we argued in the paper, the holographic zero density system is automatically a system with energy not conserved for the charge carriers. To encode energy dissipations at finite density, we can as well mimic the way that momentum dissipations are introduced, such as the Q lattice \cite{Donos:2013eha} or massive gravity constructions \cite{Vegh:2013sk}. It is possible to combine all the momentum, energy and charge dissipations holographically to test the formula in this work and we would like to consider this in future work.

%%%%%%%%%%%%%%%%%%%%%%%%%%%%%%%
\begin{figure}[h]
\begin{center}
\begin{tabular}{cc}
\includegraphics[width=0.7\textwidth]{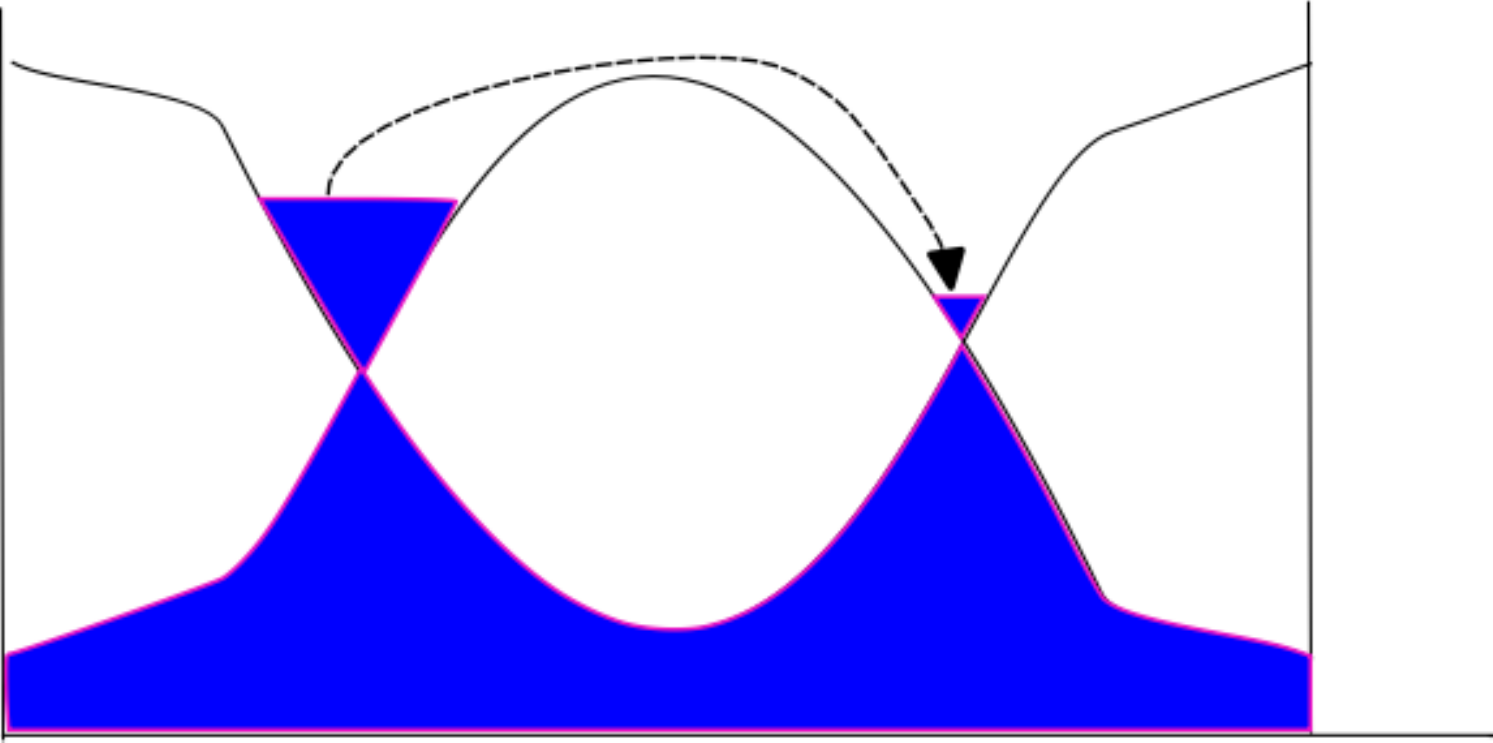}
\end{tabular}
\end{center}
\caption{\small Schematic depiction of an inter valley scattering event. Such an event will lead to axial charge relaxation. But if the
two Weyl cones are at different chemical potentials (as they are in parallel external electric and magnetic fields) inter valley
scattering will also lead to energy relaxation since $\delta \epsilon \approx \mu_5 \delta \rho_5$.}
\label{intervalley}
\end{figure}
%%%%%%%%%%%%%%%%%%%%%%%%%%%%%%%%%%%

Finally we would like to point out that in the context of Weyl metals inter-valley scattering does indeed lead to energy  relaxation.
A schematic picture of an intervalley scattering event is shown in figure \ref{intervalley}. In such an event an electron from the
Fermi surface of the left-handed Weyl cone is scattered into the right-handed Weyl cone. In the presence of parallel electric and
magnetic fields the local Fermi energies in the two Weyl cones will be shifted due to the injection of axial charge via the axial
anomaly $d\rho_5 / dt = 1/(2\pi^2) \vec E .\vec B$. The difference in the local Fermi energies can be encoded in an axial chemical
potential. Since an intervalley scattering event changes the axial charge $\rho_5 = \rho_L - \rho_R$ this is accompanied by a cost
in energy of the form $\delta \epsilon = \mu_5 \delta \rho_5$. This explains qualitatively why the energy dissipation is present
in our hydrodynamic considerations. It would be interesting to include this effect also in the kinetic theory based on a simply
collision term of axial charge relaxation along the lines in \cite{sonspivak}.

%%%%%%%%%%%%%%%%%%%%%%%%%%%%%%%%%
\section*{Appendix: Holographic Hall conductivity in the probe limit}
%%%%%%%%%%%%%%%%%%%%%%%%%%%%%%%%%
In this appendix we compute the transverse magnetoconductivity and Hall conductivity in the holographic probe limit for one  $U(1)$ case. We consider the fluctuations $\delta A_x=a_x e^{-i\omega t},\delta A_y=a_y e^{-i\omega t}.$ The equations of motion are 
\begin{eqnarray}
a_x''-\frac{2u}{1-u^2}a_x'+\bigg[\frac{\omega^2}{4r_0^2 u(1-u^2)^2}a_x-\frac{4i\omega \alpha (8\alpha B)A_t}{2r_0^4(1-u^2)^2}a_y\bigg]&=&0\,,\nonumber\\
a_y''-\frac{2u}{1-u^2}a_y'+\bigg[\frac{\omega^2}{4r_0^2 u(1-u^2)^2}a_y+\frac{4i\omega \alpha (8\alpha B)A_t}{2r_0^4(1-u^2)^2}a_x\bigg]&=&0\,.
\end{eqnarray}
Define $a_{\pm}=a_x\pm i a_y, $ we have 
\begin{equation}\label{eomHall}
a_\pm''-\frac{2u}{1-u^2}a_\pm'+\bigg[\frac{\omega^2}{4r_0^2 u(1-u^2)^2}\mp\frac{4\omega \alpha (8\alpha B)A_t}{2r_0^4(1-u^2)^2}\bigg]a_\pm=0\,.
\end{equation}
Following \cite{Hartnoll:2007ai}, redefine $a_\pm= (1-u^2)^{-\frac{i\omega}{4r_0}}\big(a_\pm^{(0)}+\omega a_\pm^{(1)}+\dots\big)$ where $a_\pm^{(0)}, a_\pm^{(1)}$ are regular at the horizon. Expanding the equation (\ref{eomHall}) according to $\omega$, at the zeroth order we have 
\begin{equation}
(a^{(0)}_\pm)''-\frac{2u}{1-u^2}(a_\pm^{(0)})'=0.
\end{equation}
Thus $a_\pm^{(0)}(u)=c_0^{\pm}.$
At the first order we have 
\begin{equation}
\big[(1-u^2)(a_\pm^{(1)})'\big]'+\bigg(\frac{i}{2r_0}\mp\frac{16\alpha^2 B A_t}{r_0^4(1-u^2)}\bigg)c_{\pm}^{0}=0.
\end{equation}
We obtain
\begin{eqnarray}
\big(a_\pm^{(1)}\big)'&=&-\frac{c_{\pm}^{0}}{1-u^2}\int_1^u dx\bigg(\frac{i}{2r_0}\mp\frac{16\alpha^2 B A_t}{r_0^4(1-x^2)}\bigg)\nonumber\\
&=&\frac{c_{\pm}^{0}}{1-u^2}\bigg(\frac{i(1-u)}{2r_0}\pm\frac{1}{B}\big[A_t'[u]-A_t'[1]\big]\bigg).
\end{eqnarray}
Following \cite{Hartnoll:2007ip}, the DC conductivity can be computed 
\begin{equation}
\mp i\sigma_{xy}+\sigma_{xx}=\lim_{\omega\to 0}\frac{2r_0^2 \omega \big(a^{(1)}_\pm\big)'(0)}{i\omega a_\pm(0)}\,.
\end{equation}
We have 
\begin{equation}\label{eq:streda}
\sigma_{xx}=\pi T, ~~~\sigma_{xy}=-\frac{\rho}{B}-\frac{2r_0^2 A_t'[1]}{B}=-\frac{\rho-\rho_h}{B}
\end{equation}
where $\rho_h$ is the charge density carried by the horizon. %of the same system with $B=0$ and fixing $\mu$. 

%%%%%%%%%%%%%%%%%%%%%%%%%%%%%%%%%%%%%%%%%
\section*{Acknowledgments}
%%%%%%%%%%%%%%%%%%%%%%%%%%%%%%%%%%%%%%%%%
We thank A.~Cortijo, J.~Gauntlett, V.~Jacobs, A.~Jimenez-Alba, H.~Liu, L.~Melgar, S.~Sachdev, H.~Stoof, M.~Vozmediano,  J.~Zaanen for discussions.
This work was supported in part
 by the Spanish MINECO's ``Centro de Excelencia Severo Ochoa" Programme under grant SEV-2012-0249 and Plan Nacional de
Altas Energ\'\i as FPA2009-07890.


\begin{thebibliography}{99}

%\cite{Fukushima:2008xe}
\bibitem{Fukushima:2008xe}
  K.~Fukushima, D.~E.~Kharzeev and H.~J.~Warringa,
  ``The Chiral Magnetic Effect,''
  Phys.\ Rev.\ D {\bf 78} (2008) 074033
  [arXiv:0808.3382 [hep-ph]].
  %%CITATION = ARXIV:0808.3382;%%
  %453 citations counted in INSPIRE as of 16 Oct 2014

\bibitem{Kharzeev:2007jp}
  D.~E.~Kharzeev, L.~D.~McLerran and H.~J.~Warringa,
  ``The Effects of topological charge change in heavy ion collisions: 'Event by event P and CP violation',''
  Nucl.\ Phys.\ A {\bf 803} (2008) 227
  [arXiv:0711.0950 [hep-ph]].
  %%CITATION = ARXIV:0711.0950;%%
  %549 citations counted in INSPIRE as of 16 Oct 2014

\bibitem{qi-review}
P. Hosur, X. L. Qi,~
``Recent developments in transport phenomena in Weyl semimetals,"
Comptes Rendus Physique Vol. 14, Issues 9-10, November-December 2013, Pages 857-870, 
[arXiv:1309.4464 [cond-mat.str-el]].

%\cite{Kaminski:2014jda}
\bibitem{Kaminski:2014jda}
  M.~Kaminski, C.~F.~Uhlemann, M.~Bleicher and J.~Schaffner-Bielich,
  ``Anomalous hydrodynamics kicks neutron stars,''
  arXiv:1410.3833 [nucl-th].
  %%CITATION = ARXIV:1410.3833;%%

%\cite{Adler:1969gk}
\bibitem{Adler:1969gk}
  S.~L.~Adler,
  ``Axial vector vertex in spinor electrodynamics,''
  Phys.\ Rev.\  {\bf 177} (1969) 2426.
  %%CITATION = PHRVA,177,2426;%%
  %2954 citations counted in INSPIRE as of 16 Oct 2014

%\cite{Bell:1969ts}
\bibitem{Bell:1969ts}
  J.~S.~Bell and R.~Jackiw,
  ``A PCAC puzzle: $\pi^0 \rightarrow  \gamma \gamma$ in the $\sigma$ model,''
  Nuovo Cim.\ A {\bf 60} (1969) 47.
  %%CITATION = NUCIA,A60,47;%%
  %2434 citations counted in INSPIRE as of 16 Oct 2014


%\cite{Son:2009tf}
\bibitem{Son:2009tf}
  D.~T.~Son and P.~Surowka,
  ``Hydrodynamics with Triangle Anomalies,''
  Phys.\ Rev.\ Lett.\  {\bf 103}, 191601 (2009)
  [arXiv:0906.5044 [hep-th]].
  %%CITATION = ARXIV:0906.5044;%%


\bibitem{Erdmenger:2008rm}
  J.~Erdmenger, M.~Haack, M.~Kaminski and A.~Yarom,
  ``Fluid dynamics of R-charged black holes,''
  JHEP {\bf 0901} (2009) 055
  [arXiv:0809.2488 [hep-th]].
  %%CITATION = ARXIV:0809.2488;%%

\bibitem{Banerjee:2008th}
  N.~Banerjee, J.~Bhattacharya, S.~Bhattacharyya, S.~Dutta, R.~Loganayagam and
P.~Surowka,
  ``Hydrodynamics from charged black branes,''
  JHEP {\bf 1101} (2011) 094
  [arXiv:0809.2596 [hep-th]].
  %%CITATION = ARXIV:0809.2596;%%

\bibitem{Neiman:2010zi}
  Y.~Neiman and Y.~Oz,
  ``Relativistic Hydrodynamics with General Anomalous Charges,''
  JHEP {\bf 1103} (2011) 023
  [arXiv:1011.5107 [hep-th]].
  %%CITATION = ARXIV:1011.5107;%%


\bibitem{Landsteiner:2011cp}
  K.~Landsteiner, E.~Megias and F.~Pena-Benitez,
  ``Gravitational Anomaly and Transport,''
  Phys.\ Rev.\ Lett.\  {\bf 107} (2011) 021601
  [arXiv:1103.5006 [hep-ph]].
  %%CITATION = ARXIV:1103.5006;%%

\bibitem{Landsteiner:2011iq}
  K.~Landsteiner, E.~Megias, L.~Melgar and F.~Pena-Benitez,
  ``Holographic Gravitational Anomaly and Chiral Vortical Effect,''
  JHEP {\bf 1109} (2011) 121
  [arXiv:1107.0368 [hep-th]].
  %%CITATION = ARXIV:1107.0368;%%

\bibitem{Jensen:2012jy}
  K.~Jensen,
  ``Triangle Anomalies, Thermodynamics, and Hydrodynamics,''
  Phys.\ Rev.\ D {\bf 85}, 125017 (2012)
  [arXiv:1203.3599 [hep-th]].
  %%CITATION = ARXIV:1203.3599;%%

\bibitem{Jensen:2012kj}
  K.~Jensen, R.~Loganayagam and A.~Yarom,
  ``Thermodynamics, gravitational anomalies and cones,''
  arXiv:1207.5824 [hep-th].
  %%CITATION = ARXIV:1207.5824;%%

%\cite{Golkar:2012kb}
\bibitem{Golkar:2012kb}
  S.~Golkar and D.~T.~Son,
  ``Non-Renormalization of the Chiral Vortical Effect Coefficient,''
  arXiv:1207.5806 [hep-th].
  %%CITATION = ARXIV:1207.5806;%%
  %30 citations counted in INSPIRE as of 24 Oct 2014

%\cite{Hou:2012xg}
\bibitem{Hou:2012xg}
  D.~F.~Hou, H.~Liu and H.~c.~Ren,
  ``A Possible Higher Order Correction to the Vortical Conductivity in a Gauge Field Plasma,''
  Phys.\ Rev.\ D {\bf 86} (2012) 121703
  [arXiv:1210.0969 [hep-th]].
  %%CITATION = ARXIV:1210.0969;%%
  %22 citations counted in INSPIRE as of 24 Oct 2014

\bibitem{Banerjee:2012iz}
  N.~Banerjee, J.~Bhattacharya, S.~Bhattacharyya, S.~Jain, S.~Minwalla and T.~Sharma,
  ``Constraints on Fluid Dynamics from Equilibrium Partition Functions,''
  JHEP {\bf 1209}, 046 (2012)
  [arXiv:1203.3544 [hep-th]].
  %%CITATION = ARXIV:1203.3544;%%

\bibitem{Kharzeev:2013ffa}
  D.~E.~Kharzeev,
  ``The Chiral Magnetic Effect and Anomaly-Induced Transport,''
  Prog.\ Part.\ Nucl.\ Phys.\  {\bf 75}, 133 (2014)
  [arXiv:1312.3348 [hep-ph]].
  %%CITATION = ARXIV:1312.3348;%%


%\cite{Nielsen:1983rb}
\bibitem{Nielsen:1983rb} 
  H.~B.~Nielsen and M.~Ninomiya,
  ``Adler-Bell-Jackiw Anomaly And Weyl Fermions In Crystal,''
  Phys.\ Lett.\ B {\bf 130}, 389 (1983).
  %%CITATION = PHLTA,B130,389;%%

 \bibitem{sonspivak} 
 D. T. Son and B. Z. Spivak,    
``Chiral Anomaly and Classical Negative Magnetoresistance of Weyl Metals,"
Phys. Rev. B. 88.104412 (2013)
[arXiv:1206.1627 [cond-mat.mes-hall]].


 \bibitem{gorbar}
E. V. Gorbar, V. A. Miransky, and I. A. Shovkovy,
``Chiral anomaly, dimensional reduction, and magnetoresistivity of Weyl and Dirac semimetals,"
Phys. Rev. B 89, 085126 (2014),
[arXiv:1312.0027 [cond-mat.mes-hall]]

\bibitem{theorem-prb}
G. H. Wannier,
``Theorem on the Magnetoconductivity of Metals,"
Phys. Rev. B {\bf 5}, 3836 (1972).

\bibitem{Lifschytz:2009si}
  G.~Lifschytz and M.~Lippert,
  ``Anomalous conductivity in holographic QCD,''
  Phys.\ Rev.\ D {\bf 80}, 066005 (2009)
  [arXiv:0904.4772 [hep-th]].
  %%CITATION = ARXIV:0904.4772;%%
  %42 citations counted in INSPIRE as of 22 Oct 2014}

\bibitem{Buividovich:2010tn}
  P.~V.~Buividovich, M.~N.~Chernodub, D.~E.~Kharzeev, T.~Kalaydzhyan, E.~V.~Luschevskaya and M.~I.~Polikarpov,
  ``Magnetic-Field-Induced insulator-conductor transition in SU(2) quenched lattice gauge theory,''
  Phys.\ Rev.\ Lett.\  {\bf 105} (2010) 132001
  [arXiv:1003.2180 [hep-lat]]. %\cite{Buividovich:2010qe}

\bibitem{Buividovich:2010qe} 
  P.~V.~Buividovich and M.~I.~Polikarpov,
  ``Quark mass dependence of the vacuum electric conductivity induced by the magnetic field in SU(2) lattice gluodynamics,''
  Phys.\ Rev.\ D {\bf 83}, 094508 (2011)
  [arXiv:1011.3001 [hep-lat]].
  %%CITATION = ARXIV:1011.3001;%%
  %%CITATION = ARXIV:1003.2180;%%
  %85 citations counted in INSPIRE as of 24 Oct 2014

\bibitem{Chernodub:2010qx}
  M.~N.~Chernodub,
  ``Superconductivity of QCD vacuum in strong magnetic field,''
  Phys.\ Rev.\ D {\bf 82}, 085011 (2010)
  [arXiv:1008.1055 [hep-ph]].
  %%CITATION = ARXIV:1008.1055;%%
  %114 citations counted in INSPIRE as of 24 Oct 2014


\bibitem{weyl-exp}
H.-J. Kim, K.-S. Kim, J. F. Wang, M. Sasaki, N. Satoh, A. Ohnishi, M. Kitaura, M. Yang, and L. Li,
``Dirac vs. Weyl in topological insulators: Adler-Bell-Jackiw anomaly in transport phenomena,"
Phys. Rev. Lett. 111, 246603 (2013),
[arXiv:1307.6990 [cond-mat.str-el]].


\bibitem{kim-boltzman}
K.-S Kim, H.-J. Kim, M. Sasaki, 
``Boltzmann-equation approach to anomalous transport in a Weyl metal,"
Phys. Rev. B 89, 195137 (2014), 
[arXiv:1402.4240 [cond-mat.str-el]]. 

\bibitem{kim-review}
K.-S Kim, H.-J. Kim, M. Sasaki
``Anomalous transport phenomena in Weyl metal beyond the Drude model for Landau's Fermi liquids,"
[arXiv:1407.3056 [cond-mat.mes-hall]]


%\cite{Jimenez-Alba:2014iia}
\bibitem{Jimenez-Alba:2014iia} 
  A.~Jimenez-Alba, K.~Landsteiner and L.~Melgar,
  ``Anomalous Magneto Response and the St\"uckelberg Axion in Holography,''
  arXiv:1407.8162 [hep-th].
  %%CITATION = ARXIV:1407.8162;%%

\bibitem{linearresponse}
  L. P. Kadanoff and P. C. Martin, 
``Hydrodynamic equations and correlation functions,"
 Annals of Physics (N.Y.) 24, 419 (1963)

\bibitem{nernsteffect}
S.~Hartnoll, P.~Kovtun, M.~Mueller and S.~Sachdev, 
``Theory of the Nernst effect near quantum phase transitions in condensed matter, and in dyonic black holes,"
Phys.\ Rev.\ B {\bf 76}, 144502 (2007)
[arXiv:0706.3215 [cond-mat.str-el]].


%\cite{Hansen:2008tq}
\bibitem{Hansen:2008tq} 
  J.~Hansen and P.~Kraus,
 ``Nonlinear Magnetohydrodynamics from Gravity,''
  JHEP {\bf 0904}, 048 (2009)
  [arXiv:0811.3468 [hep-th]].
  %%CITATION = ARXIV:0811.3468;%%

\bibitem{geracie-son}
M.~Geracie and D. T. Son, 
``Hydrodynamics on the lowest Landau level,"
[arXiv:1408.6843 [cond-mat.mes-hall]].


%\cite{Landsteiner:2012kd}
\bibitem{Landsteiner:2012kd}
  K.~Landsteiner, E.~Megias and F.~Pena-Benitez,
  ``Anomalous Transport from Kubo Formulae,''
  Lect.\ Notes Phys.\  {\bf 871}, 433 (2013)
  [arXiv:1207.5808 [hep-th]].
  %%CITATION = ARXIV:1207.5808;%%

\bibitem{elias} 
B. S. Kim, E. Kiritsis, C. Panagopoulos, 
``Holographic quantum criticality and strange metal transport,"
New J. Phys. 14: 043045, 2012, 
[arXiv:1012.3464 [cond-mat.str-el]].

%\cite{Gynther:2010ed}
\bibitem{Gynther:2010ed}
  A.~Gynther, K.~Landsteiner, F.~Pena-Benitez and A.~Rebhan,
 ``Holographic Anomalous Conductivities and the Chiral Magnetic Effect,''
  JHEP {\bf 1102}, 110 (2011)
  [arXiv:1005.2587 [hep-th]].
  %%CITATION = ARXIV:1005.2587;%%

%\cite{Horowitz:2008bn}
\bibitem{Horowitz:2008bn}
  G.~T.~Horowitz and M.~M.~Roberts,
  ``Holographic Superconductors with Various Condensates,''
  Phys.\ Rev.\ D {\bf 78}, 126008 (2008)
  [arXiv:0810.1077 [hep-th]].
  %%CITATION = ARXIV:0810.1077;%%


%\cite{Kovtun:2008kx}
\bibitem{Kovtun:2008kx} 
  P.~Kovtun and A.~Ritz,
 ``Universal conductivity and central charges,''
  Phys.\ Rev.\ D {\bf 78}, 066009 (2008)
  [arXiv:0806.0110 [hep-th]].
  %%CITATION = ARXIV:0806.0110;%%

%\cite{Karch:2007pd}
\bibitem{Karch:2007pd} 
  A.~Karch and A.~O'Bannon,
  ``Metallic AdS/CFT,''
  JHEP {\bf 0709}, 024 (2007)
  [arXiv:0705.3870 [hep-th]].
  %%CITATION = ARXIV:0705.3870;%%

%\cite{Donos:2014cya}
\bibitem{Donos:2014cya} 
  A.~Donos and J.~P.~Gauntlett,
  ``Thermoelectric DC conductivities from black hole horizons,''
  arXiv:1406.4742 [hep-th].
  %%CITATION = ARXIV:1406.4742;%%

%\cite{Gorbar:2009bm}
\bibitem{Gorbar:2009bm}
  E.~V.~Gorbar, V.~A.~Miransky and I.~A.~Shovkovy,
  ``Chiral asymmetry of the Fermi surface in dense relativistic matter in a magnetic field,''
  Phys.\ Rev.\ C {\bf 80}, 032801 (2009)
  [arXiv:0904.2164 [hep-ph]].
  %%CITATION = ARXIV:0904.2164;%%
  %36 citations counted in INSPIRE as of 23 Oct 2014

%\cite{Gorbar:2013qsa}
\bibitem{Gorbar:2013qsa}
  E.~V.~Gorbar, V.~A.~Miransky and I.~A.~Shovkovy,
  ``Engineering Weyl nodes in Dirac semimetals by a magnetic field,''
  Phys.\ Rev.\ B {\bf 88} (2013) 165105
  [arXiv:1307.6230 [cond-mat.mes-hall]].
  %%CITATION = ARXIV:1307.6230;%%
  %8 citations counted in INSPIRE as of 23 Oct 2014

%\cite{Sadofyev:2010pr}
\bibitem{Sadofyev:2010pr} 
  A.~V.~Sadofyev and M.~V.~Isachenkov,
  ``The Chiral magnetic effect in hydrodynamical approach,''
  Phys.\ Lett.\ B {\bf 697}, 404 (2011)
  [arXiv:1010.1550 [hep-th]].
  %%CITATION = ARXIV:1010.1550;%%
  
%\cite{Kalaydzhyan:2011vx}
\bibitem{Kalaydzhyan:2011vx} 
  T.~Kalaydzhyan and I.~Kirsch,
  ``Fluid/gravity model for the chiral magnetic effect,''
  Phys.\ Rev.\ Lett.\  {\bf 106}, 211601 (2011)
  [arXiv:1102.4334 [hep-th]].
  %%CITATION = ARXIV:1102.4334;%%  



%\cite{Horowitz:2012ky}
\bibitem{Horowitz:2012ky} 
  G.~T.~Horowitz, J.~E.~Santos and D.~Tong,
  ``Optical Conductivity with Holographic Lattices,''
  JHEP {\bf 1207}, 168 (2012)
  [arXiv:1204.0519 [hep-th]].
  %%CITATION = ARXIV:1204.0519;%%
%\cite{Liu:2012tr}

\bibitem{Liu:2012tr} 
  Y.~Liu, K.~Schalm, Y.~W.~Sun and J.~Zaanen,
  ``Lattice Potentials and Fermions in Holographic non Fermi-Liquids: Hybridizing Local Quantum Criticality,''
  JHEP {\bf 1210}, 036 (2012)
  [arXiv:1205.5227 [hep-th]].
  %%CITATION = ARXIV:1205.5227;%%  

 %\cite{Donos:2012js}
\bibitem{Donos:2012js} 
  A.~Donos and S.~A.~Hartnoll,
  ``Interaction-driven localization in holography,''
  Nature Phys.\  {\bf 9}, 649 (2013)
  [arXiv:1212.2998].
  %%CITATION = ARXIV:1212.2998;%% 

%\cite{Donos:2013eha}
\bibitem{Donos:2013eha} 
  A.~Donos and J.~P.~Gauntlett,
  ``Holographic Q-lattices,''
  JHEP {\bf 1404}, 040 (2014)
  [arXiv:1311.3292 [hep-th]].
  %%CITATION = ARXIV:1311.3292;%%

%\cite{Andrade:2013gsa}
\bibitem{Andrade:2013gsa} 
  T.~Andrade and B.~Withers,
  ``A simple holographic model of momentum relaxation,''
  JHEP {\bf 1405}, 101 (2014)
  [arXiv:1311.5157 [hep-th]].
  %%CITATION = ARXIV:1311.5157;%%  

%\cite{Vegh:2013sk}
\bibitem{Vegh:2013sk} 
  D.~Vegh,
  ``Holography without translational symmetry,''
  arXiv:1301.0537 [hep-th].
  %%CITATION = ARXIV:1301.0537;%%
%\cite{Davison:2013jba}

\bibitem{Davison:2013jba} 
  R.~A.~Davison,
  ``Momentum relaxation in holographic massive gravity,''
  Phys.\ Rev.\ D {\bf 88}, 086003 (2013)
  [arXiv:1306.5792 [hep-th]].
  %%CITATION = ARXIV:1306.5792;%%
%\cite{Davison:2013txa}

\bibitem{Davison:2013txa} 
  R.~A.~Davison, K.~Schalm and J.~Zaanen,
  ``Holographic duality and the resistivity of strange metals,''
  Phys.\ Rev.\ B {\bf 89}, 245116 (2014)
  [arXiv:1311.2451 [hep-th]].
  %%CITATION = ARXIV:1311.2451;%%


%\cite{Gursoy:2014ela}
\bibitem{Gursoy:2014ela} 
  U.~Gursoy and A.~Jansen,
  ``(Non)renormalization of Anomalous Conductivities and Holography,''
  arXiv:1407.3282 [hep-th].
  %%CITATION = ARXIV:1407.3282;%%
  

\bibitem{inprogress}
A.~Jimenez-Alba, K.~Landsteiner, Y.~Liu and Y.-W. Sun, in progress.


  %\cite{Hartnoll:2007ai}
\bibitem{Hartnoll:2007ai} 
  S.~A.~Hartnoll and P.~Kovtun,
  ``Hall conductivity from dyonic black holes,''
  Phys.\ Rev.\ D {\bf 76}, 066001 (2007)
  [arXiv:0704.1160 [hep-th]].
  %%CITATION = ARXIV:0704.1160;%%
  
  %\cite{Hartnoll:2007ip}
\bibitem{Hartnoll:2007ip} 
  S.~A.~Hartnoll and C.~P.~Herzog,
  ``Ohm's Law at strong coupling: S duality and the cyclotron resonance,''
  Phys.\ Rev.\ D {\bf 76}, 106012 (2007)
  [arXiv:0706.3228 [hep-th]].
  %%CITATION = ARXIV:0706.3228;%%


\end{thebibliography}
\end{document}